\documentclass[twocolumn]{aastex62} 
\usepackage{amsmath,amstext}
\usepackage[figure,figure*]{hypcap}
\usepackage{subfigure,graphicx}
\usepackage{comment}
\usepackage{lmodern}
\usepackage[T1]{fontenc}

\newcommand{\mjup}{\ifmmode {M_{\rm Jup}}\else${M_{\rm Jup}}$\fi}

\newcommand{\msun}{\ifmmode {M_{\odot}}\else${M_{\odot}}$\fi}
\newcommand{\rsun}{\ifmmode {R_{\odot}}\else${R_{\odot}}$\fi}
\newcommand{\Msun}{\ifmmode {M_{\odot}}\else${M_{\odot}}$\fi}
\newcommand{\Rsun}{\ifmmode {R_{\odot}}\else${R_{\odot}}$\fi}
\newcommand{\Lsun}{\ifmmode {L_{\odot}}\else${L_{\odot}}$\fi}
\newcommand{\lapprox }{{\lower0.8ex\hbox{$\buildrel <\over\sim$}}}
\newcommand{\gapprox }{{\lower0.8ex\hbox{$\buildrel >\over\sim$}}}
\def\amin{\ifmmode^{\prime}\else$^{\prime}$\fi}
\def\asec{\ifmmode^{\prime\prime}\else$^{\prime\prime}$\fi}

\newcommand{\degree}{\ifmmode {^\circ}\else$^\circ$\fi}

\newcommand{\Ro}{\ifmmode {R_o}\else$R_o\ $\fi}
\newcommand{\lha}{\ifmmode {L_{H\alpha}/L_{bol}}\else$L_{H\alpha}/L_{bol}$\ \fi}

\newcommand{\caiihk}{\ion{Ca}{2} H \& K}

\newcommand{\rphk}{\ensuremath{R'_{\rm HK}}}
\newcommand{\lrphk}{\ensuremath{\log{\rphk}}}

\newcommand{\teff}{\ensuremath{T_{\mbox{\scriptsize eff}}}}
\newcommand{\logg}{\ensuremath{\log g}}
\newcommand{\prot}{\ensuremath{P_{rot}}} 
\newcommand{\porb}{\ensuremath{P_{orb}}} 
\newcommand{\vsini}{\ensuremath{v \sin i}}
\newcommand{\kms}{\ensuremath{\mbox{km s}^{-1}}}

\newcommand{\mas}{\ensuremath{\mbox{mas yr}^{-1}}}

\newcommand{\gbr}{(\ensuremath{G_{\rm BP} - G_{\rm RP}})}

\shorttitle{{\it K2} Periods for Hyads}
\shortauthors{Douglas et al.}

\begin{document}

\title{K2 rotation periods for low-mass Hyads and a quantitative comparison of the distribution of slow rotators in the Hyades and Praesepe}

\correspondingauthor{S.~T.~Douglas}
\email{StephanieTDouglas@gmail.com}

\author[0000-0001-7371-2832]{S. T. Douglas}
\altaffiliation{NSF Astronomy and Astrophysics Postdoctoral Fellow}
\affiliation{Center for Astrophysics | Harvard\ \&\ Smithsonian, 60 Garden St, Cambridge, MA 02138, USA}

\author[0000-0002-2792-134X]{J. L. Curtis}
\altaffiliation{NSF Astronomy and Astrophysics Postdoctoral Fellow}
\affiliation{Department of Astronomy, Columbia University, 550 West 120th Street, New York, NY 10027, USA}

\author[0000-0001-7077-3664]{M. A. Ag{\"u}eros}
\affiliation{Department of Astronomy, Columbia University, 550 West 120th Street, New York, NY 10027, USA}

\author[0000-0002-1617-8917]{P. A. Cargile}
\affiliation{Center for Astrophysics | Harvard\ \&\ Smithsonian, 60 Garden St, Cambridge, MA 02138, USA}

\author[0000-0002-9873-1471]{J. M. Brewer}
\affiliation{Department of Astronomy, Yale University, 52 Hillhouse Avenue, New Haven, CT 06511, USA}
\affiliation{Department of Astronomy, Columbia University, 550 West 120th Street, New York, NY 10027, USA}

\author{S. Meibom}
\affiliation{Center for Astrophysics | Harvard\ \&\ Smithsonian, 60 Garden St, Cambridge, MA 02138, USA}

\author[0000-0002-8444-3436]{T. Jansen}
\affiliation{Department of Astronomy, Columbia University, 550 West 120th Street, New York, NY 10027, USA}

\begin{abstract}
We analyze {\it K2} light curves for 132 low-mass ($1\ \gapprox\ M_*\  \gapprox\ 0.1$~\Msun) members of the 600--800~Myr-old Hyades cluster and measure rotation periods (\prot) for 116 of these stars. These include 93 stars with no prior \prot\ measurement; the total number of Hyads with known \prot\ is now 232. We then combine literature binary data with \textit{Gaia} DR2 photometry and astrometry to select single star sequences in the Hyades and its roughly coeval Praesepe open cluster,
and derive a new reddening value of $A_V = 0.035$$\pm$$0.011$ for Praesepe.
Comparing the effective temperature--\prot\ distributions for the Hyades and Praesepe, we find that solar-type Hyads rotate, on average, 0.4~d slower than their Praesepe counterparts.
This \prot\ difference indicates that the Hyades is slightly older than Praesepe: we apply a new gyrochronology model tuned with Praesepe 
and the Sun, and find an age difference between the two clusters of 57~Myr. 
However, this \prot\ difference decreases and eventually disappears for lower-mass stars. 
This provides further evidence for stalling in the rotational evolution of these stars, and highlights the need for 
more detailed analysis of angular-momentum evolution for stars of different masses and ages.
\end{abstract}


\section{Introduction}\label{intro}

The Hyades and Praesepe open clusters are benchmarks for 
determining the dependence of stellar rotation on age.
The Hyades was one of the first open clusters for which photometric rotation periods ($P_{rot}$) were measured for low-mass stars \citep[$\lapprox$1~\Msun;][]{radick1987,radick1995}. 
The two clusters are sufficiently nearby such that many photometric $P_{rot}$ across the full FGKM mass range have now been measured for both 
from ground- and space-based photometric monitoring
\citep[e.g.,][]{agueros11, delorme2011, hartman2011, douglas2014,douglas2016, douglas2017, rebull2017}.

Empirical efforts to establish the functional form of the rotation-age relation, sometimes referred to as gyrochronology \citep{barnes2003}, have commonly assumed that the mass dependence can be separated from 
the age dependence, such that 
$\prot(M_\star, t)  = f(M_\star) \times g(t)$.
This was famously proposed by \citet{skumanich72}, who found that solar-type stars spin down as $\prot \propto t^n$, where the braking index $n\approx0.5$. 
\citet{barnes2003,barnes2007} accounted for the dependence on mass by adopting photometric color as its observational proxy, 
then fit coefficients for a simple 
analytic function from observations of rotators with a range of masses in young nearby clusters.
The resulting model implied that lower-mass stars spin down more rapidly than their solar-type counterparts. Later authors \citep[e.g.,][]{mamajek2008,meibom2009,angus2015} have adjusted the coefficients and braking index, but otherwise have assumed the same functional form as \citeauthor{barnes2007}.
However, an examination of the figures in \citet{barnes2003} shows that this fixed relation between $P_{rot}$, $t$, and color is insufficient to describe stellar spin-down for stars with a range of masses.

More recent \prot\ measurements for G and K dwarfs in open clusters have shown that $P_rot$ evolution cannot be described by separating the mass and age dependence.
Using the Skumanich relation, \citet{meibom2011} tested whether Hyades rotators could be spun up to match the observed distribution of \prot\ in M34, which is 220 Myr old. 
These authors determined that while the distribution of spun-up solar-type Hyads did match that of their younger cousins, spinning up Hyades K dwarfs by the same factor resulted in these stars having faster $P_{rot}$ than those observed in M34.
Comparing $P_{rot}$  measured for GKM stars in various open clusters from 100~Myr to 1~Gyr
leads to a similar conclusion: \citeauthor{skumanich72}-like spin-down works well for solar-type stars, but K dwarfs spin down more slowly 
\citep{meibom2011,cargile2014,Agueros2018}. 

Furthermore, while re-tuning the coefficients for the \citet{barnes2007} gyrochronology equation, \citet{angus2015} could not simultaneously fit Praesepe and the Hyades. 
When including Praesepe, these authors' fit resulted in a multi-modal distribution for their color singularity term, which controls the downturn toward rapid rotation for bluer/hotter/more massive stars (which have thinner convective envelopes, resulting in relatively weaker magnetic dynamos and braking efficiency). 
This is additional evidence that the shape of the slow-rotator sequence can vary from cluster to cluster.

A complication in using the Hyades and Praesepe for calibrating gyrochronology is that their absolute and relative ages, usually determined from isochrones, are still debated (see Table~\ref{table:ages} for examples of ages derived for the two clusters). Most authors agree that the clusters are either coeval, or that the Hyades is slightly older, 
and that their ages range from $\approx$600 to $\approx$800 Myr. 
But the disagreements among these ages do not provide much hope that we can successfully calibrate gyrochronology using isochronal cluster ages. And it creates confusion for gyrochronology studies: some authors separate the two clusters when comparing data to theoretical models \citep[][]{brown2014,matt2015,garraffo2018}, while others combine them \citep[][]{reiners2012,angus2015}.

\begin{deluxetable}{lcc}[!t]
\tablewidth{0pt}
\tablecaption{Literature Ages for the Hyades and Praesepe \label{table:ages}}
\tablehead{
\colhead{} & \colhead{Hyades}  & \colhead{Praesepe} \\
\colhead{} & \colhead{(Myr)}  & \colhead{(Myr)}
}
\startdata
\citet{perryman1998} & 625$\pm$50 & \nodata	 \\
\citet{fossati2008}  & \nodata & $590^{+150}_{-120}$ \\ 
\citet{brandt2015-2} &  750$\pm$100& \nodata  \\ 
\citet{brandt2015-1} & $790\pm60$ & $790\pm60$ \\ 
\citet{david2015}\tablenotemark{a} & $827^{+10}_{-15}$ & \nodata \\
 & $764^{+16}_{-17}$ & \nodata \\
\citet{Choi2016} & \nodata & 630 \\ 
\citet{Cummings2017} &  635$\pm$25 & 670$\pm$25 \\
\citet{Cummings2018}\tablenotemark{b} &  705$\pm$25 & 700$\pm$25 \\
 &  705$\pm$25 & 685$\pm$25 \\
\citet{gossage2018}\tablenotemark{c} & $676^{+67}_{-11}$ & $617^{+40}_{-10}$ \\ 
 & $741^{+55}_{-12}$ & $617^{+17}_{-15}$ \\ 
& $676^{+13}_{-30}$ & $589^{+13}_{-26}$ \\ 
 & $589^{+29}_{-11}$ & $617^{+14}_{-13}$ \\ 
\citet{DR2HRD} & 794 & 708 \\ %
\enddata
\tablenotemark{a}{\citet{david2015} fit two different isochrone models; we give both results from their summed PDF analysis in log age space.}
\tablenotemark{b}{\citet{Cummings2018} fit two different isochrone models; we list both results.}
\tablenotemark{c}{\citet{gossage2018} fit models with different rotation parameterizations to both ($B$,$V$) and ($J$,$K_s$) photometry. We give the results from fitting the model with a free rotation parameter and the model with a fixed rotation parameter but a spread in rotation to both color-magnitude diagrams.}
\end{deluxetable}

Our goal is to compare the shapes of the slow-rotator sequences in the Hyades and Praesepe and to determine whether they can be combined into a single benchmark sample for gyrochronology. 
\citet{delorme2011} carried out a similar anaysis. These authors compared $P_{rot}$ distributions in the Hyades, Praesepe, and Coma Berenices (thought to be of similar age).  Using a simple linear fit to the color--period relation, they found the Hyades to be $\approx$50~Myr older than the other two clusters.
However, \citet{delorme2011} did not have access to \textit{Gaia} data for membership, nor did they have the wealth of new $P_{rot}$ measurements enabled by \textit{K2}'s observations of the Hyades and Praesepe.
We use updated catalogs of rotators in both clusters to carry out our analysis.

We describe our membership and archival $P_{rot}$ catalogs for the two clusters in Section~\ref{data} before deriving masses ($M_*$) and effective temperatures (\teff) for these stars in Section~\ref{prop}. In Section~\ref{binaries}, we identify binaries among our \textit{K2} targets. Binary companions can impact the rotational evolution of a star, and therefore confuse interpretation of the mass-period distribution of a cluster. We then present new $P_{rot}$ measurements for 116 Hyads from \textit{K2} Campaign 13 in Section~\ref{prot}. Finally, we derive single-star sequences in both clusters using the second \textit{Gaia} data release
\citep[DR2;][]{GAIADR2},
obtain a new reddening value for Praesepe, and derive a differential gyrochronological age for the Hyades in Section~\ref{hypra}.
We discuss our results and their potential implications for calibrating angular momentum evolution in Section~\ref{res}, and conclude in Section~\ref{concl}.

\section{Existing Data}\label{data}
\subsection{Hyades Membership and Rotation Catalog}\label{cats}
As in \cite{douglas2014,douglas2016}, we use the \citet{roser2011} and \citet{goldman2013} catalogs as the basis for our work.
To these we add 13 stars identified using reduced proper motions and parallaxes from \textit{Hipparcos}, bringing us to 786 total Hyads.
Since archival data for the Hyades are generally of high quality, and since our pre-\textit{Gaia} catalog was used to select our \textit{K2} Campaign 4 and 13 targets (Guest Observer proposals 4095 and 13064), we do not attempt to update the full cluster membership list using \textit{Gaia} DR2.

Furthermore, since our sample consists of variable stars and includes probable binaries, these stars will have increased photometric variability and possibly also high astrometric excess noise. This variability and excess noise will impact the availability of the \textit{Gaia} data, as well as the determination of appropriate quality cuts.
Indeed, 188 stars in our original catalog do not pass the quality cuts recommended by the \textit{Gaia} Collaboration \citep{DR2HRD}, and $>$80 of these are confirmed or candidate binaries.  

\begin{figure}[t]
\centerline{\includegraphics[width=\columnwidth]{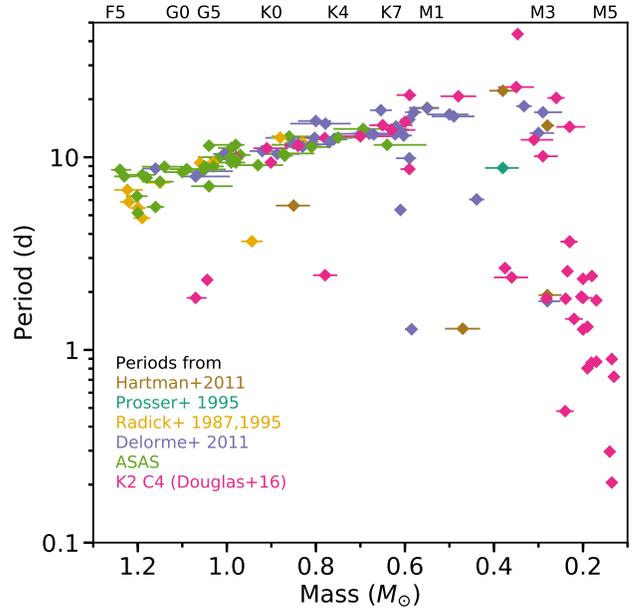}}
\caption{Mass-period distribution for Hyads with $P_{rot}$ measurements in the literature. The color indicates the source of the $P_{rot}$.
We also include the uncertainties on $M_*$, which are dominated by distance uncertainties even in the \textit{Gaia} DR2 era. The error bars only represent systematic uncertainties from our mass calculation, and do not reflect, e.g., systematics in the model or excess $K$-band flux due to an unresolved companion.
}
\label{fig:periodmass_lit}
\end{figure}

In \cite{douglas2014,douglas2016}, we assembled $P_{rot}$ measurements for Hyads from \citet{radick1987, radick1995}; \citet{prosser1995};  \citet{delorme2011}; \citet{hartman2011}; and from an analysis of All Sky Automated Survey \citep[ASAS;][]{ASAS} data (A.~Kundert \& P.~Cargile, private communication, 2014)\footnote{In \citet{douglas2014} we cited these $P_{rot}$ as Kundert et al.\ in prep, and in \citet{douglas2016} as Cargile et al.\ in prep. These periods were measured by A.~Kundert as an undergraduate while being supervised by co-author P.~Cargile. The paper was never completed, however, and additional ASAS data have become available for Hyades members in the last few years. We therefore give the existing ASAS $P_{rot}$ measurements in Table~\ref{tab:k2}, but further details will be provided in a later paper, where we will re-analyze the expanded ASAS data set. Since we find that ground-based $P_{rot}$ generally, and ASAS $P_{rot}$ specifically, are consistent with \textit{K2} periods, we feel justified in continuing to include the current ASAS periods in our analysis.} into a catalog of 102 rotators.
We then added 37 new $P_{rot}$ from our analysis of \textit{K2} Campaign 4 data in \citet{douglas2016}, bringing the total number of known Hyades rotators to 139.
With a few exceptions, these surveys generally measure consistent $P_{rot}$; for details, see
\cite{douglas2014,douglas2016}. 
The mass-period relationship for these 139 Hyads is shown in Figure~\ref{fig:periodmass_lit}.

In the second half of this paper, we consider only single, slowly rotating Hyads, and we use \textit{Gaia} data to select these stars. We match our \citet{douglas2016} Hyades catalog 
to \textit{Gaia} DR2, and select the nearest neighbor. We then check this match by computing synthetic \textit{Gaia} $G$ magnitudes from UCAC $r$, $i$ magnitudes \citep[][]{zacharias2010}, SDSS $r$, $i$ \citep[][]{alam2015},
2MASS $J$, $K$ \citep[][]{2mass}, and/or Tycho2 $B$, $V$ (as given in 2MASS). We require that at least one of these synthetic magnitudes match the measured \textit{Gaia} $G$ value to within 1 standard deviation ($\sigma$) for optical photometry or to within 2$\sigma$ for 2MASS photometry.  Of the 786 stars in our catalog, only 10 fail this test: three stars lack photometry to compute synthetic $G$ magnitudes, two lack \textit{Gaia} counterparts, and five fail the $G$ magnitude test. However, none of those 10 stars has a measured $P_{rot}$ or is a \textit{K2} target, so they do not impact our analysis and are excluded from all tables.

\subsection{Praesepe Membership and Rotation Catalog}
We continue to use the \citet{douglas2017} Praesepe membership catalog, which is based primarily on \citet{adam2007}.
Our catalog includes 1130 cluster members with $P_{mem}\geq50\%$ from \citet{adam2007}, supplemented by 39 previously cataloged members too bright to be identified by those authors.
We assign these bright stars $P_{mem}=100$\%.

In \cite{douglas2014,douglas2017}, we gathered $P_{rot}$ measurements for Praesepe members from \citet{scholz2007}, \citet{scholz2011}, \citet{delorme2011}, \citet{agueros11}, and \citet{kovacs2014}.
We combined these literature values with 677 $P_{rot}$ derived from our \textit{K2} Campaign 5 data; in total, our catalog includes $P_{rot}$ data for 743 Praesepe members.

We match this list of Praesepe rotators to \textit{Gaia} DR2 and again select the nearest neighbor.
Only three rotators in our catalog lack a DR2 match within 0\farcm1: EPIC 211970974 and EPIC 211907026 are both rapidly rotating M dwarfs, and EPIC 211954582 is overluminous by $-$1.18 mag, which suggests that it might be a triple system.
Since our analysis focuses on single, slowly rotating stars, the lack of a DR2 match in these three cases does not affect this work.

Five additional stars were mismatched when searching for the nearest neighbor, but in each case, another star was found within 0\farcm1 with photometry consistent with our target: \\
\textit{Gaia} DR2 661314466963687808 (EPIC 211971468),
\textit{Gaia} DR2 659680072990872704 (EPIC 211903302),
\textit{Gaia} DR2 661355934869899648 (EPIC 211983811),
\textit{Gaia} DR2 663055371825360000 (EPIC 211981509),
\textit{Gaia} DR2 661312267940341632 (EPIC 211966619). 

\section{Derived stellar properties}\label{prop}
\subsection{Stellar Masses}\label{masses}
As in previous work, we estimate stellar masses by linearly interpolating between the $M_K$ and $M_\star$ points given by \citet{adam2007}, who list
$M_\star$ and spectral energy distributions (SEDs) for B8-L0 stars.

We calculate distances ($D$) to individual stars using \textit{Gaia} DR2 or \textit{Hipparcos} \citep{perryman1998} parallaxes, or the secular parallaxes from \citet{roser2011} or \citet{goldman2013}.
For stars passing the \textit{Gaia} quality cuts, we use \textit{Gaia} parallaxes.
For the remaining stars, we use \textit{Hipparcos} parallaxes or secular parallaxes.
We then use these distances to compute $M_K$.

\begin{deluxetable*}{llrrrcccl}[t]
\tablewidth{0pt}
\tabletypesize{\scriptsize}
\tablecaption{Confirmed and candidate multiple systems among \textbf{Hyades K2 targets and members} with Measured $P_{rot}$ \label{tab:bin}}
\tablehead{
\colhead{[RSP2011]\tablenotemark{a}} & \colhead{HIP} & \colhead{2MASS J} & \colhead{EPIC} & \colhead{D16} & \colhead{Updated} & \colhead{Gaia} & \colhead{Conf?} & \colhead{Ref}\\
\colhead{} & \colhead{} & \colhead{} & \colhead{} & \colhead{Cand?} & \colhead{Cand?} & \colhead{Cand?} & \colhead{} & \colhead{}
}
\startdata
323 & \nodata & 04260584+1531275 & \nodata & N & N & N & Y & \citet{patience1998}; R.~Stefanik (priv.~comm.)\\
293 & 20577 & 04242831+1653103 & \nodata & Y & Y & N & Y & \citet{douglas2014, patience1998, kopytova2016} \\
360 & 20899 & 04284827+1717079 & \nodata & N & N & N & Y & \citet{mason2001} \\
329 & 20719 & 04262460+1651118 & \nodata & Y & Y & N & Y & \citet{douglas2014, mermilliod2009} \\
330 & 20741 & 04264010+1644488 & \nodata & N & N & N & Y &  \citet{morzinski2011} \\
530 & 22203 & 04463036+1528194 & \nodata & N & N & N & Y & \citet{morzinski2011}; R.~Stefanik (priv.~comm.)
\enddata
\tablecomments{This table is available in its entirety in machine-readable form.}
\tablenotetext{a}{Index in the \citet{roser2011} catalog}
\end{deluxetable*}

We also propagate the $m_K$ 
and $D$ uncertainties for each star to determine the $M_*$ uncertainties, $\sigma_{M_*}$.
The uncertainties are typically small, on the order of a few percent.
In our previous work, a few stars had large uncertainties in $D$, which led to large mass uncertainties. The improved parallaxes from \textit{Gaia} have remedied this.
Our stated $\sigma_{M_*}$ are only the systematic uncertainties resulting from our calculation and the chosen model;
they do not take into account other sources of uncertainty, such as our choice of model or $K$-band excesses due to a binary companion.

\subsection{Effective Temperatures}\label{temp}

In Section~\ref{hypra}, we also compare the two clusters' $P_{rot}$-\teff\ relations.
For solar-type stars with
$4700 < \teff < 6700$ K,
we derive an empirical color--\teff\ relation
using a \textit{Gaia} DR2 match to the
California Planet Survey catalog \citep{Brewer2016}.
For warmer stars, we supplement this with Hyades members
from \citet{DR2HRD} with \teff\ from DR2/Apsis
\citep{DR2prop}.
For cooler stars,
we combine the benchmark K and M dwarfs from
\citet{Mann2015} and \citet{Boyajian2012}.
That sample only reaches $\teff > 3056$ K,
so we also adopt the \citet{Rabus2019} $M_G$--\teff\ relation for stars with $2600 < \teff < 4000$ K. At $\teff = 4000$ K, our color--\teff\ relation predicts a value only 9 K different from the \citet{Rabus2019} formula when using our fit to the Hyades main-sequence to convert between color and absolute magnitude.

\section{Binary Identification} \label{binaries}

We search binaries among known rotators in the Hyades because they can bias our analysis of the $P_{rot}$ distribution.
Binary companions may exert tidal or other physical effects on the primary star \citep[e.g.,][]{meibom2005, meibom2007,zahn2008,douglas2016,douglas2017}.
In addition, when two (or more) stars are blended in a given image, the second star may dilute the rotational signal and/or add flux that will cause us to overestimate $L_{bol}$ and $M_*$.
These effects can cause stars to be misplaced in the mass--period plane, leading us to misidentify trends or transitions in the period distribution.
Finally, short-period binaries are susceptible to tidal interactions, which can cause atypical angular momentum evolution.
Binaries with orbital periods under $\sim$10~days might be circularized and locked, but others with orbital periods up to 30~days could still be affected.
We therefore wish to identify as many binary systems as possible among our Hyades {\it K2} targets. We denote all confirmed and candidate binaries in our analysis, and provide a brief overview of our binary identification methods below.
For more details, see   \citet{douglas2016,douglas2017}.

\begin{enumerate}
\item \textit{visual identification:} We examine a co-added \textit{K2} image, a Digital Sky Survey (DSS) red image, and a 2MASS \citep{cutri2003cat} $K$-band image of each target to look for neighboring stars (see Figure~\ref{fig:k2lc}).
We use a flag of ``Y'' for yes, ``M'' for maybe, and ``N'' for no to indicate whether the target and a neighbor have blended point spread functions (PSFs) on the \textit{K2} chip.
Stars flagged as ``Y'' are labeled candidate binaries; we find 38 such targets, or 29\% of stars with \textit{K2} $P_{rot}$.

By searching 12 regions of the nearby sky in \textit{Gaia} DR2, we find the rate of chance alignments with $G\le20$ mag stars within 10\arcsec\ to be $\approx$6--58\%. We find a range in potential contamination rates because the Hyades is so large on the sky: part of the cluster sits close to the Galactic Plane, but it also extends well away from the Plane. At typical Hyades distances, 10\arcsec\ corresponds to $\approx$400--550~AU; it is possible that all of the blends we identify are chance alignments, or that up to 23\% of Hyads have a companion within $\approx$400--550~AU \citep[for comparison, we determined that $\approx$10\% of Praesepe members likely have a bound companion within 10\arcsec, or $10^3$--$10^4$~AU;][]{douglas2017}.
For consistency with our previous work, we continue to label probable blends as candidate binaries. 

\item \textit{photometric identification:} As in previous work, we identify candidate unresolved binaries that are overluminous for their color.
In \citet{douglas2014,douglas2016,douglas2017}, we selected binaries that were overluminous in a $r\prime$~vs~($r\prime- K_{S}$) color-magnitude diagram (CMD), using \textit{Hipparcos} \citet{perryman1998} parallaxes or secular parallaxes from \citet{roser2011} and \citet{goldman2013}.
As in Section~\ref{masses}, we now update the $r\prime$~vs~($r\prime- K_{S}$) selection using \textit{Gaia}~DR2 parallaxes when the data passes the quality cuts defined in \citet{DR2HRD}.
We also select new photometric candidate binaries using \textit{Gaia} DR2 photometry, discussed further in Section~\ref{phot_selection}.
This method is biased towards binaries with equal masses, and we are certainly missing candidate binaries with lower mass ratios.
Our binary selections are shown in Figure~\ref{fig:binarycmd}; in Section~\ref{prot} we flag all photometric candidate binaries, but in Section~\ref{hypra} we reject only candidates selected from \textit{Gaia} photometry.

\begin{figure*}[t]
\begin{center}
\includegraphics[  width=3.4in]{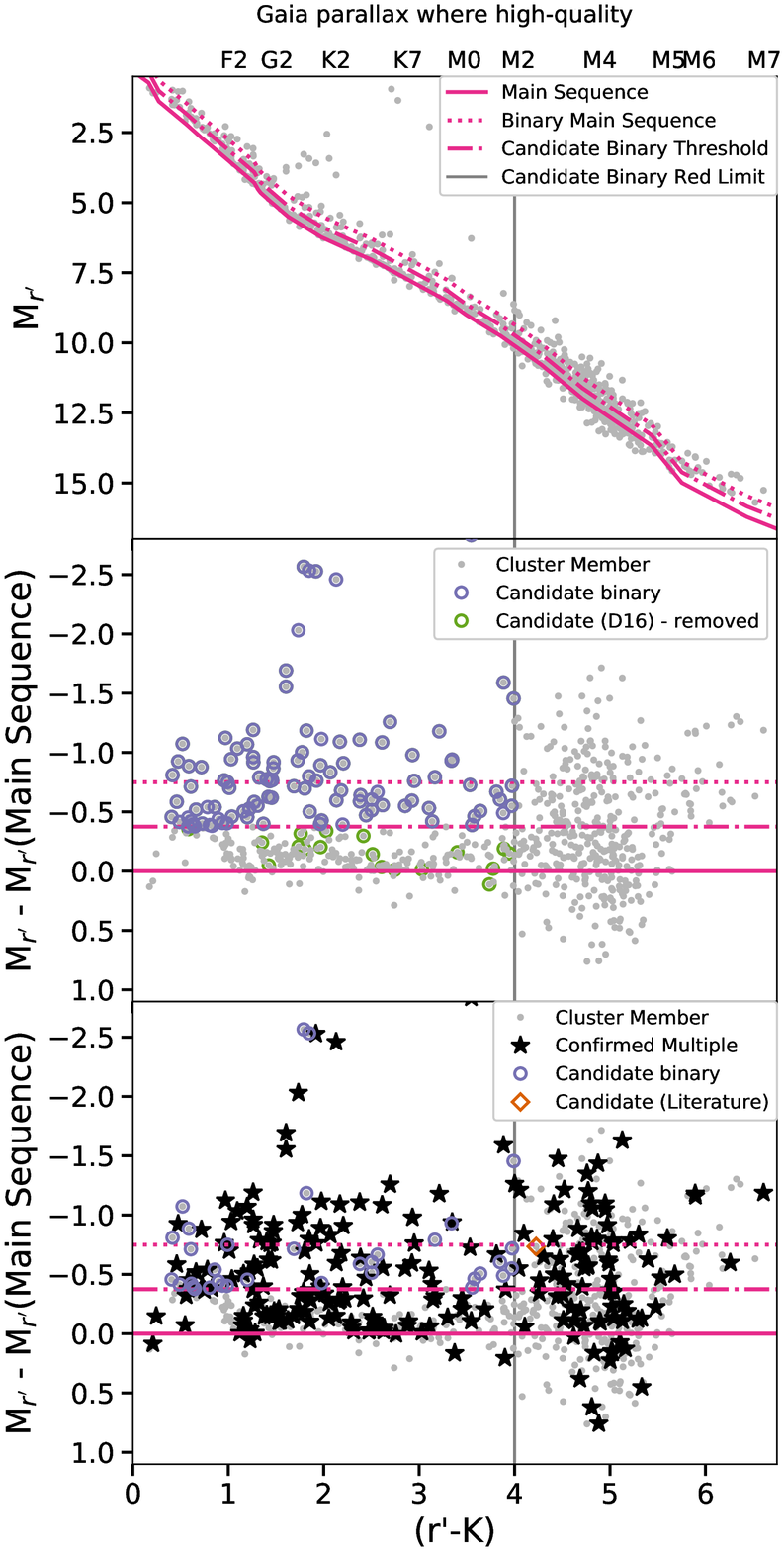}
\includegraphics[  width=3.4in]{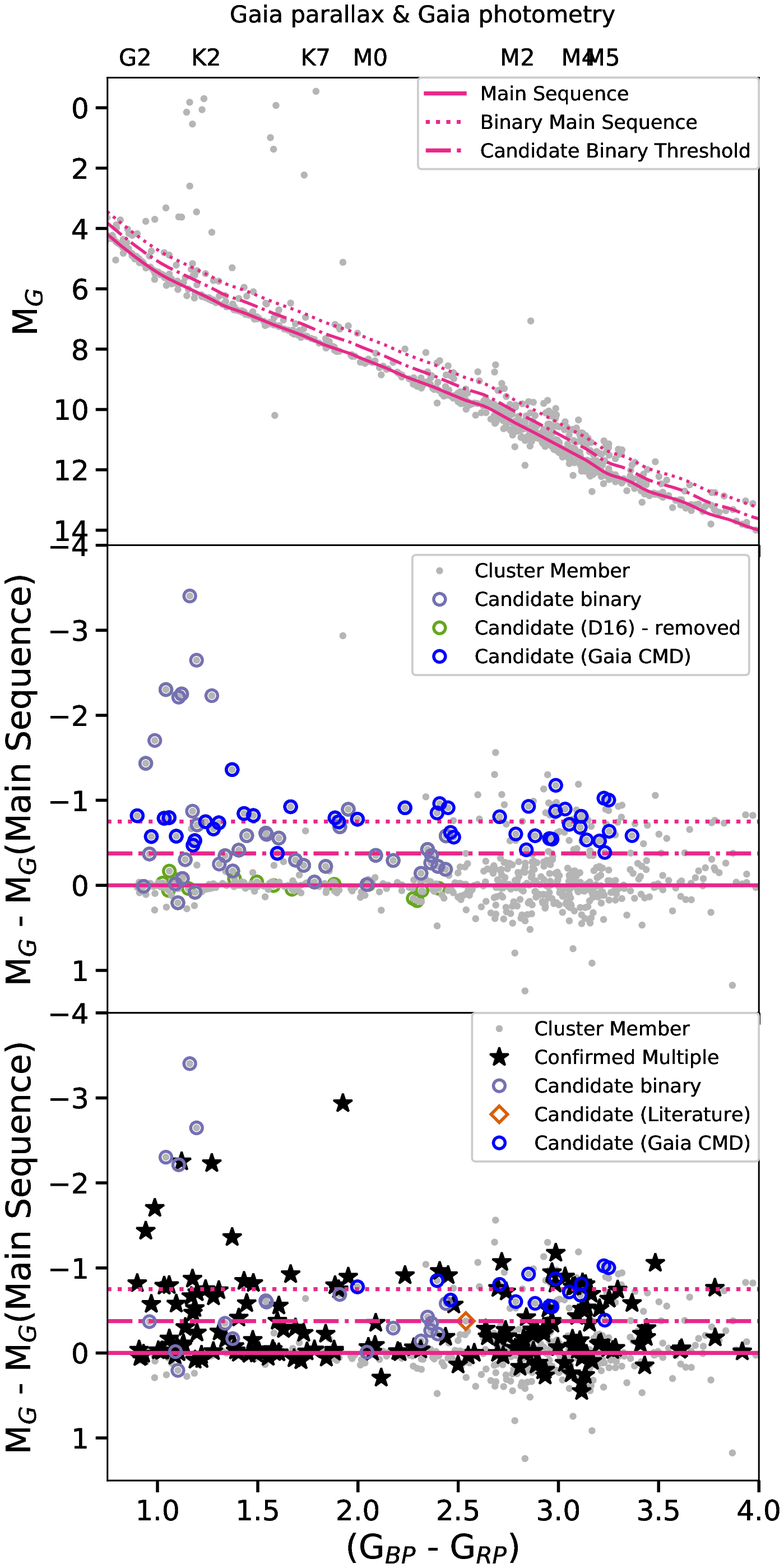}
  \caption{Demonstration of our photometric binary selection process. The left column shows the ($r^{\prime}-K_s$) color we used in previous work, but using {\it Gaia} parallaxes where available to determine $M_{r^{\prime}}$. The right column shows one of the {\it Gaia} color-magnitude diagrams (CMDs) we used to update our candidate binary list for this work.
  \textit{Top}---CMD with our selected main sequence (solid line) and binary cuts overlaid: the dotted line shows the nominal binary main sequence, and the dot-dashed line gives the minimum magnitude above which we consider a star to be a candidate binary. In previous work (left) we use the model SEDs assembled by \citet{adam2007}, and in this work (right) we use a polynomial fit to the Hyades main sequence.
  \textit{Middle}---residuals between observed and expected absolute magnitudes; the horizontal lines are the same thresholds given above. Candidates identified in \citet{douglas2016} are shown in in green, and new candidates identified from Gaia DR2 photometry are given in blue. It is clear that the improved Gaia parallaxes have removed some of our previously identified candidate binaries.
  \textit{Bottom}---the same as the middle panel, but now confirmed multiples (black stars) and literature candidates (orange diamonds) are also shown. While our photometric selection is useful for identifying additional candidates, there are still many confirmed binaries that show no photometric offset from the main sequence.
  }
    \label{fig:binarycmd}
\end{center}\end{figure*}


\item \textit{multiperiodic K2 stars:} In binaries where the components have roughly equal brightness, variability from both stars can appear in the \textit{K2} light curve.
However, we may also detect two $P_{rot}$ and/or an obvious beat pattern when a single star exhibits differential rotation.
As discussed in Section~\ref{prot}, we assume that the two periods come from different components of a binary if the periods are different by $>$20\%.
This cutoff is based on the maximum period separation for differentially rotating spot groups on the Sun. We find multiple $P_{rot}$, indicating probable unresolved binaries, in 11 \textit{K2} targets.

\item \textit{literature identifications:} We searched the literature for Hyades binaries among known rotators and {\it K2} Campaign 4 targets in \citet{douglas2016}. We update this list with binaries among our Campaign 13 targets. We also add binaries identified or confirmed through observations with the Tillinghast Reflector Echelle Spectrograph (TRES) on the 1.5-m Tillinghast telescope at the Smithsonian Astrophysical Observatory's Fred L.~Whipple Observatory on Mt.~Hopkins, AZ (R.~Stefanik, private communication, 2018). \end{enumerate}

We consider all visual and photometric pairs, as well as multiperiodic \textit{K2} stars, to be candidate binaries in our analysis. For other literature binaries, we follow the confirmed versus candidate nomenclature used in the source paper. The resulting list of confirmed and candidate binaries is given in Table~\ref{tab:bin}. 

\section{Measuring New Hyades Rotation Periods with K2}\label{prot}

\begin{figure}[t]
\centerline{\includegraphics[width=\columnwidth]{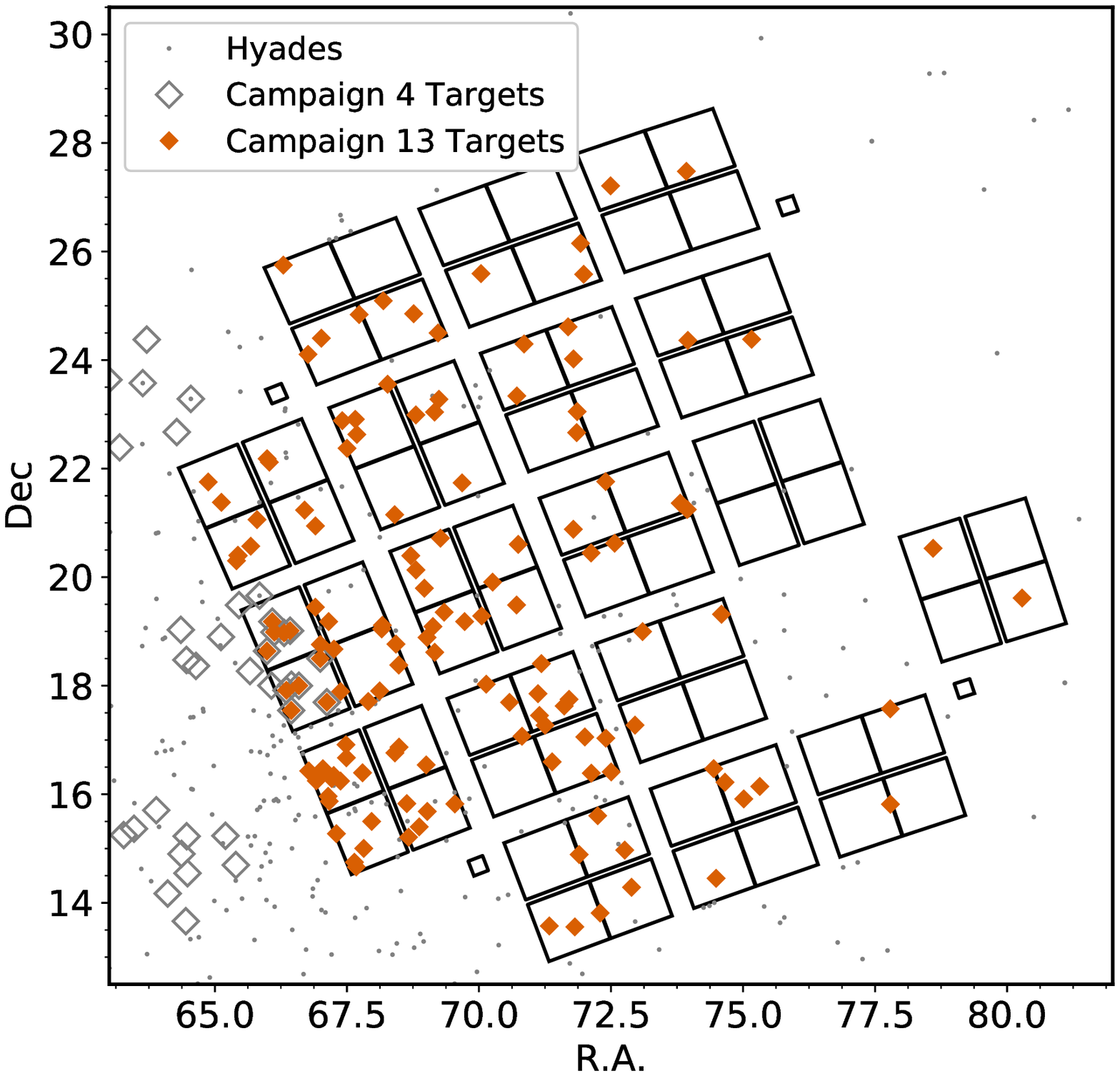}}
\caption{The {\it K2} Campaign 13 field of view, with our 132 Campaign 13 targets (orange diamonds), Campaign 4 targets (grey open diamonds), and other Hyads (grey dots).
Three of the spacecraft's detector modules are no longer functioning, but only a handful of Hyads would have fallen on these modules.
Because the cluster is so large on the sky, many targets are near the edges of the field of view, and therefore have distorted PSFs.
}
\label{fig:k2fov}
\end{figure}

\begin{deluxetable*}{lrrrrrrrrrrrrrr}[t]
\tablewidth{0pt}
\tabletypesize{\scriptsize}
\tablecaption{$P_{rot}$ measurements for Hyades stars targeted in \textit{K2} and in the literature \label{tab:k2}}
\tablehead{\colhead{[RSP2011]\tablenotemark{a}} & \colhead{EPIC} & \colhead{$P_{rot,1}$} & \colhead{$Q_1$\tablenotemark{b}} & \colhead{$P_{rot,2}$} & \colhead{$Q_2$\tablenotemark{b}} & \colhead{Multi\tablenotemark{c}} & \colhead{Blend\tablenotemark{d}} & \colhead{P\tablenotemark{e}} & \colhead{Radick} & \colhead{Prosser} & \colhead{HATnet} & \colhead{SWASP} & \colhead{ASAS} & \colhead{\textit{K2} C4 }\\
\colhead{} & \colhead{} & \colhead{(days)} & \colhead{} & \colhead{(days)} & \colhead{} & \colhead{} & \colhead{} & \colhead{} & \colhead{$P_{rot}$ (days)} & \colhead{$P_{rot}$ (days)} & \colhead{$P_{rot}$ (days)} & \colhead{$P_{rot}$ (days)} & \colhead{$P_{rot}$ (days)} & \colhead{$P_{rot}$ (days)}
}
\startdata
549 & 248045685 & 40.10 & 2 & \nodata & \nodata & N & N & \nodata & \nodata & \nodata & \nodata & \nodata & \nodata & \nodata \\
544 & 247611242 & 11.79 & 0 & \nodata & \nodata & N & Y & 3 & \nodata & \nodata & \nodata & \nodata & \nodata & \nodata \\
428 & 247369717 & 11.83 & 1 & \nodata & \nodata & M & N & K & \nodata & \nodata & \nodata & 12.69 & 13.59 & \nodata \\
571 & 246865157 & 12.22 & 0 & \nodata & \nodata & N & N & D & \nodata & \nodata & \nodata & 11.98 & \nodata & \nodata \\
362 & 210554781 & 11.51 & 0 & 3.60 & 0 & Y & N & R & 3.66 & \nodata & \nodata & \nodata & \nodata & \nodata \\
409 & 246777832 & 12.90 & 0 & \nodata & \nodata & N & N & D & \nodata & \nodata & \nodata & 13.13 & \nodata & \nodata \\
553 & 246931087 & 10.81 & 0 & \nodata & \nodata & N & Y & D & \nodata & \nodata & \nodata & 10.77 & \nodata & \nodata \\
578 & 246732310 & 12.90 & 0 & \nodata & \nodata & N & Y & D & \nodata & \nodata & \nodata & 13.14 & \nodata & \nodata \\
355 & 210651981 & 2.45 & 0 & 1.07 & 1 & M & Y & 2 & \nodata & 2.42 & \nodata & 2.42 & \nodata & 2.44 \\
658 & 246806983 & 2.61 & 0 & 14.31 & 1 & Y & N & D & \nodata & \nodata & \nodata & 14.94 & \nodata & \nodata
\enddata
\tablecomments{This table is available in its entirety in machine-readable form. }
\tablenotetext{a}{Index in the \citet{roser2011} catalog}
\tablenotetext{b}{Quality of the $P_{rot}$ detection. 0 is a high-confidence measurement, 1 is questionable, 2 is not trusted, and 3 indicates that there were no significant periodogram peaks.}
\tablenotetext{c}{Presence of multiple periods in the light curve. Y, M, and N represent ``yes'', ``maybe'', and ``no'', respectively }
\tablenotetext{d}{Presence of a blended neighbor. Y, M, and N represent ``yes'', ``maybe'', and ``no'', respectively }
\tablenotetext{e}{Flag for the $P_{rot}$ source selected. ``R'':\citet{radick1987,radick1995}, ``P'':\citet{prosser1995}, ``H'':\citet{hartman2011} (HATnet), ``D'':\citet{delorme2011} (SWASP), ``A'':ASAS, ``2'':\cite{douglas2016} (\textit{K2} Campaign 4), and ``3'':this work (\textit{K2} Campaign 13) }
\end{deluxetable*}

\subsection{K2 Data and Initial $P_{rot}$ Measurement}

{\it K2} targeted the Hyades for a second time during its Campaign 13, which lasted from 2017 Mar 08 to 2017 May 27. We analyze the resulting long-cadence data for 132 Hyads identified in Section~\ref{cats} and with {\it Kepler} magnitudes $K_p>9$~mag and $M_*<1.5$~\Msun. These limits exclude saturated stars as well as stars with radiative outer layers, which are outside of the scope of this work. The distribution of Hyades targets in {\it K2} Campaigns 4 and 13 is shown in Figure~\ref{fig:k2fov}.

We use detrended light curves generated using the {\it K2} Systematics Correction method \citep[\texttt{K2SC};][]{aigrain2016} for our analysis.
\citet{aigrain2016} developed a semi-parametric Gaussian process model to simultaneously correct for the spacecraft motion and model the stellar variability.
As discussed in \cite{douglas2017}, we find that this approach is best at removing instrumental signals and trends while leaving stellar periodic signals intact.\footnote{For more information, see \citet{aigrain2016} and the MAST high level science product page,
\url{https://archive.stsci.edu/missions/hlsp/k2sc/hlsp_k2sc_k2_llc_all_kepler_v1_readme.txt}.}
We ran the \texttt{K2SC} code on the \textit{K2} PDC pipeline light curves ourselves since the processed K2SC light curves for Campaign 13 are not yet on MAST.
We downloaded the pipeline light curves in March 2018.

We follow the same period measurement method used in \cite{douglas2017}, and only summarize it here.
We use the \citet{press1989} FFT-based Lomb-Scargle algorithm\footnote{Implemented as {\it lomb\_scargle\_fast} in the \texttt{gatspy} package; see \url{https://github.com/astroML/gatspy}.} to measure $P_{rot}$. We compute the Lomb-Scargle periodogram power for 3$\times$10$^4$ periods ranging from 0.1 to 70~d (approximately the length of the Campaign).
We also compute minimum significance thresholds for the periodogram peaks using bootstrap re-sampling, and only consider a peak to be significant if its power is greater than the minimum significance threshold for that light curve.
We take the highest significant peak as our default $P_{rot}$ value; only three of our targets show no significant periodogram peaks.

\begin{figure*}[p]
\centerline{\includegraphics[width=\textwidth]{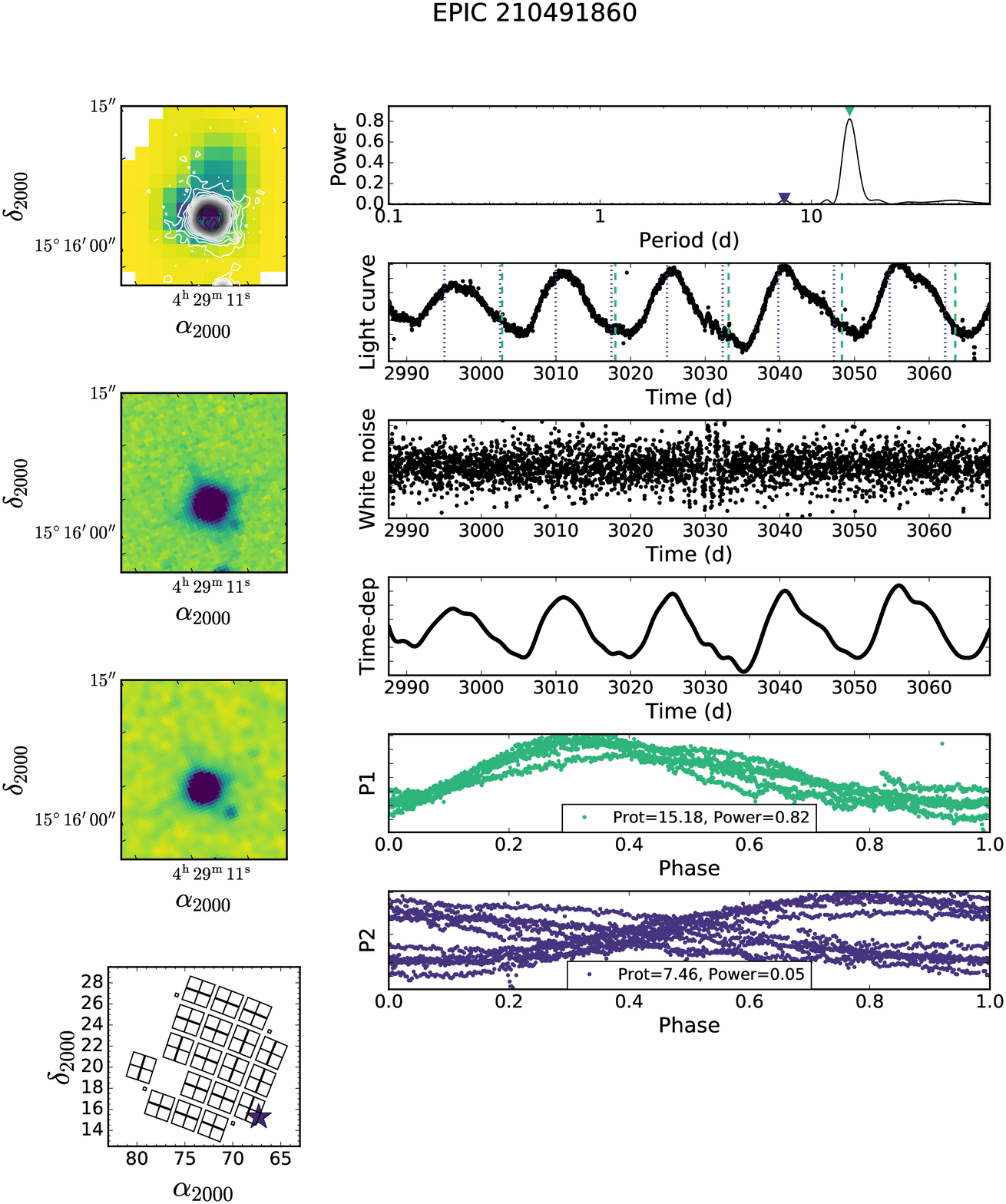}}
\caption{An example of the plots used to inspect period detections and check for neighboring stars.
{\it Left column, top to bottom}---{\it K2} pixel stamp with DSS Red image overlaid as a contour; DSS Red image rotated into the {\it K2} frame; 2MASS image rotated into the {\it K2} frame; and the target's position within the {\it K2} Campaign 13 field of view. A faint companion is visible in both the DSS and 2MASS images.
{\it Right column, top to bottom}---Lomb-Scargle periodogram with (up to) the three highest significant peaks indicated by inverted triangles; the light curve corrected for spacecraft drift; the white-noise component of the light curve; the time-dependent component; and the light curve phase-folded on (up to) the three most significant periods.
The colors of the markers indicating the peaks in the periodogram correspond to the colors of the phase-folded light curves. Slight spot evolution is apparent, and the \textit{K2SC} algorithm struggles around the middle of the campaign.
Versions of this plot for every {\it K2} target analyzed are available as an electronic figure set.}
\label{fig:k2lc}
\end{figure*}

\begin{figure}[t]
\centerline{\includegraphics[width=\columnwidth]{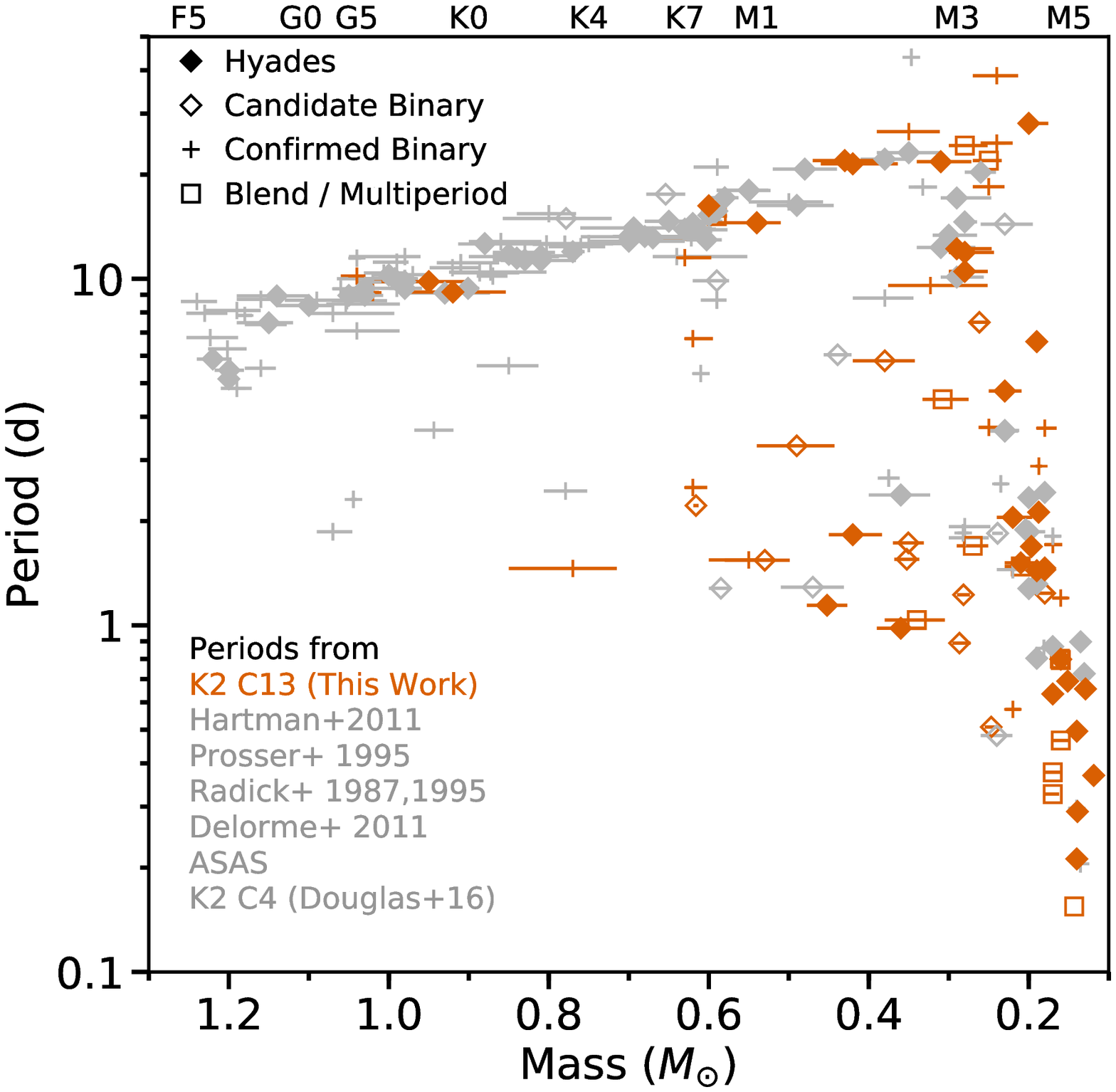}}
\caption{Hyades mass-$P_{rot}$ plane showing literature (grey) and new high-quality {\it K2} (dark orange) $P_{rot}$. 
We also mark confirmed and candidate binaries: plusses indicate confirmed binaries, open diamonds indicate photometric or spectroscopic candidate binaries, and open squares indicate {\it K2} targets with a blended neighbor or a second period in the light curve.
Approximate spectral types are indicated along the top.
All rapidly rotating stars with spectral types earlier than $\sim$K5 are confirmed binaries.
}
\label{fig:periodmass_k2}
\end{figure}

\subsection{Period Validation}\label{validation}
We employ several automated and by-eye quality checks to validate the $P_{rot}$ identified above.
We inspect each phase-folded light curve to confirm that the detected $P_{rot}$ appears astrophysical and not instrumental.
Clearly spurious detections are flagged as Q~$=2$, and questionable detections as Q~$=1$.
A Q~$=3$ flag indicates that there were no significant periodogram peaks.
Figure~3 in \citet{douglas2017} shows examples of various light curve features, and describes how we flag them.

We also plot the full light curve with vertical dashed lines at intervals corresponding to the detected $P_{rot}$, to ensure that light curve features repeat over several intervals.
Finally, we check for cases where there is a double-dip in the light curve, and the highest periodogram peak likely corresponds to half of the true $P_{rot}$. This is caused by two similar spot groups on opposite sides of the star. We then select the correct peak as the final $P_{rot}$.

Figure~\ref{fig:k2lc} shows an example of the plots we use to inspect the data; we include a figure set showing these plots for every target in our sample online.

We find 13 stars with significant periodogram peaks but no believable $P_{rot}$.
In six cases, the light curve is just noise or displays only a long trend, without any detected periodic variability.
In the remaining seven cases, there is some probable spot-induced variability, but the phase-folded light curves do not actually match up and there is no clear period.
In these cases, we are likely observing rapid spot evolution, perhaps on two stars in a binary.

For 18 other stars, the highest periodogram peak does not appear to correspond to the true $P_{rot}$. In some cases, as above, the highest periodogram peak comes from a campaign-long trend, and the true period is detected at a weaker power.
In other cases, we find a double-dip light curve with almost no difference between the central (half-period) dip and the primary (full-period) dip. In these cases, the phase-folded light curve for the longer period shows the double-dip pattern clearly, even though it is detected at a lower periodogram power.

EPIC~210741091 and EPIC~247337843 are two very interesting cases: it is hard to define a period because the spot modulation only appears in half the campaign.
For EPIC~210741091, there is initially some variability but no clear periodic signal; a V-shaped dip suggesting a single large spot \citep{bopp1973,eker1994} appears about halfway through the campaign. Nonetheless, we measure $P_{rot}=11.78$~d for this star, very close to the $P_{rot}=11.98$~d value we measured in Campaign 4.
EPIC~247337843 develops rapidly from cycle to cycle, from a slight double-dip at the beginning of the campaign to variability with no clear period by the second half.
Given this variability and parial lack of signal for both stars, we assign Q~$=1$ for their Campaign 13 $P_{rot}$.

Finally, in 11 light curves we detect two  signals with periods differing by at least 20\%. 
We consider these stars to be candidate binaries.
Several other stars exhibit two close but distinct periodogram peaks, and the light curves have obvious beat patterns.
This suggests that in these cases we are observing differential rotation of two spot groups at different latitudes.

\begin{figure}[t]
\centerline{\includegraphics[width=\columnwidth]{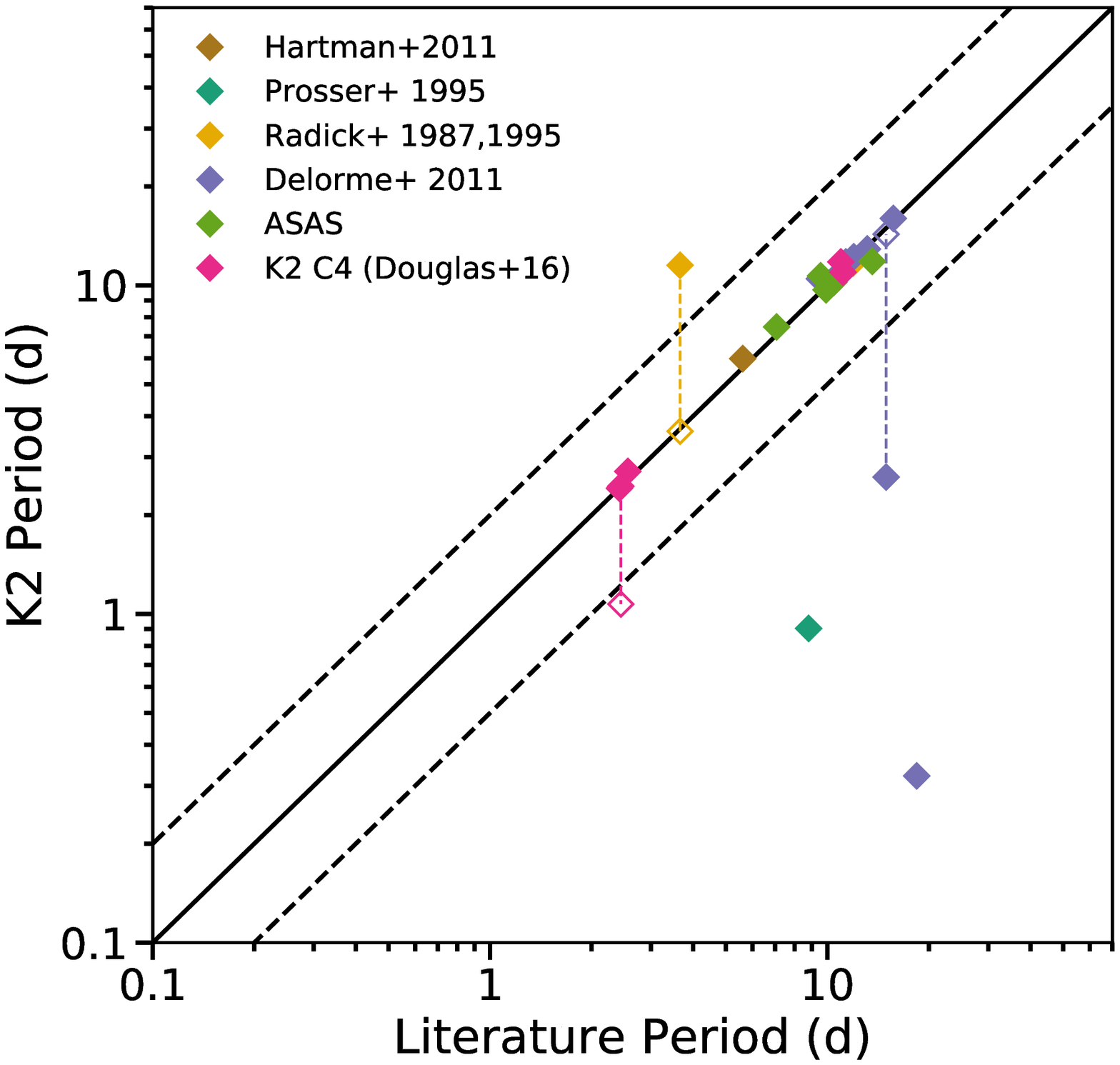}}
\caption{Comparison between the literature $P_{rot}$ and that measured from \textit{K2} Campaign 13.
Solid diamonds represent the primary period we detect in K2 Campaign 13 data, while open diamonds represent the secondary period, if any. In two cases, the literature period was detected at lower significance in the \textit{K2} light curve.
Overall, we find good agreement between \textit{K2} campaigns and between \textit{K2} and ground-based studies.
}
\label{fig:consistency}
\end{figure}

\subsection{Summary: New \textit{K2} Periods for the Hyades}
We obtain robust $P_{rot}$ measurements for 116 Hyades members, including 93 members with no prior $P_{rot}$ measurement.
The vast majority of these periods are for rapidly rotating M dwarfs, and bring the total number of Hyads with $P_{rot}$ to 232.
Our $P_{rot}$ values, flags, and analysis outputs are found in Table~\ref{tab:k2}. Our new rotation periods, along with literature values, are shown as a function of stellar mass in Figure~\ref{fig:periodmass_k2}.

Only 23 stars have $P_{rot}$ measured here and in previous studies, including five with a $P_{rot}$ measurement from \textit{K2} Campaign~4 \citep{douglas2016}.
Figure~\ref{fig:consistency} shows a comparison of the existing data with our new measurements.
In two cases (EPIC 210554781 and EPIC 246806983), the literature period is also detected as a secondary period in the \textit{K2} light curve. In two other cases (EPIC 210558541 and EPIC 246714118), we detect a short $P_{rot}$ in \textit{K2} and do not detect the longer literature $P_{rot}$ at all.
In general, however, we find that ground- and space-based $P_{rot}$ measurements agree to within 10\%, similar to our results in Praesepe \citep{douglas2017}.

\section{Comparing the Hyades and Praesepe}\label{hypra}
Based on the similarity of their color--magnitude diagrams (CMDs) and their activity, rotation, and lithium abundance data, the Hyades and Praesepe are often assumed to be coeval clusters \citep[e.g.,][]{douglas2014, Cummings2017}. Here, we test this assumption using our expanded rotator samples 
paired with the high-precision data from \textit{Gaia} DR2 for each cluster. First, we identify likely single-star members of each cluster. 
Then we apply a new gyrochronology model tuned with the Praesepe slow-rotating sequence and the Sun to infer a precise, relative, gyrochronological age for the Hyades.

\begin{figure*}\begin{center}
\includegraphics[trim=1.2cm 0cm 0.3cm 0cm, clip=True,  width=3.5in]{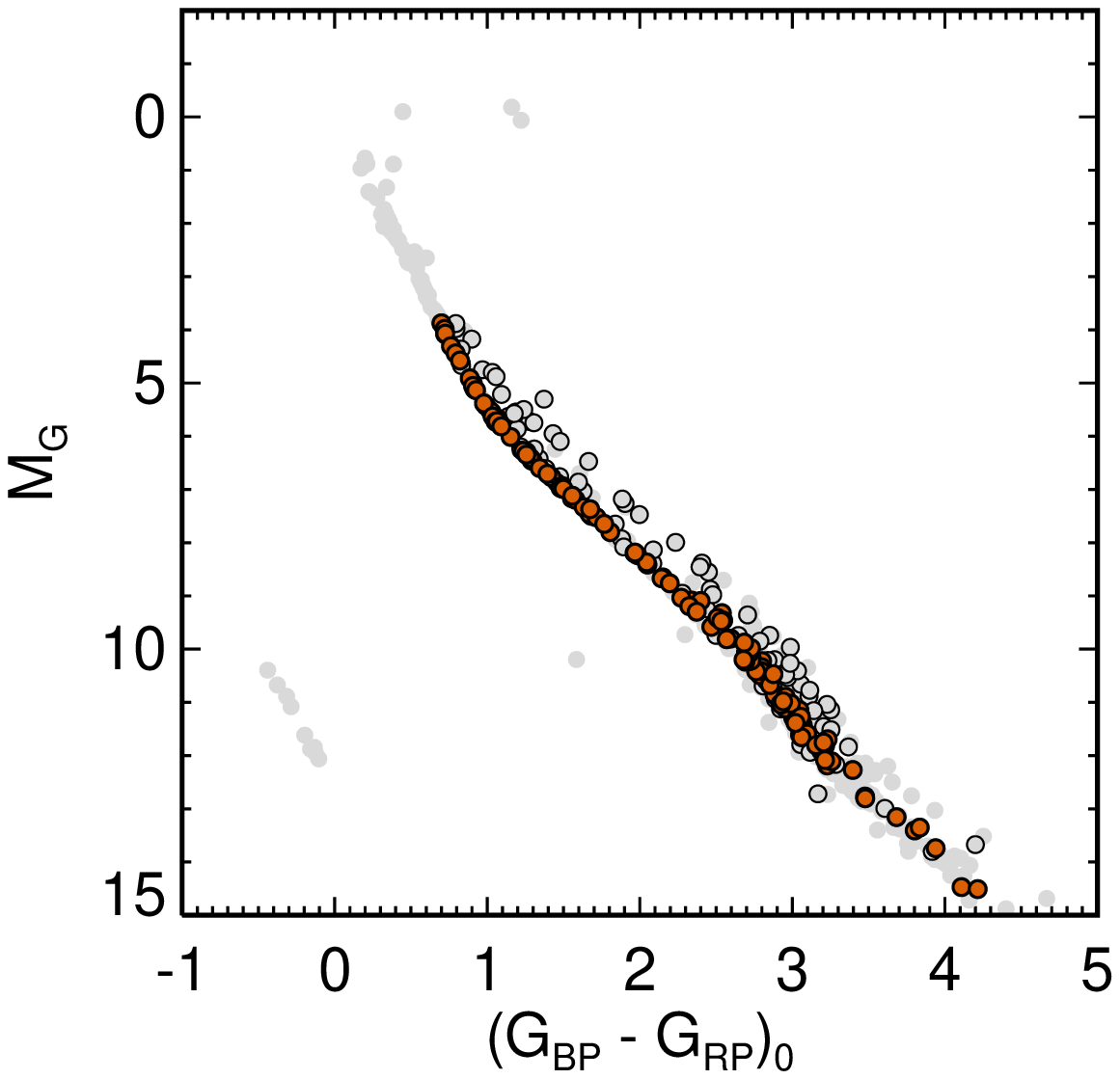}
\includegraphics[trim=1.2cm 0cm 0.3cm 0cm, clip=True,  width=3.5in]{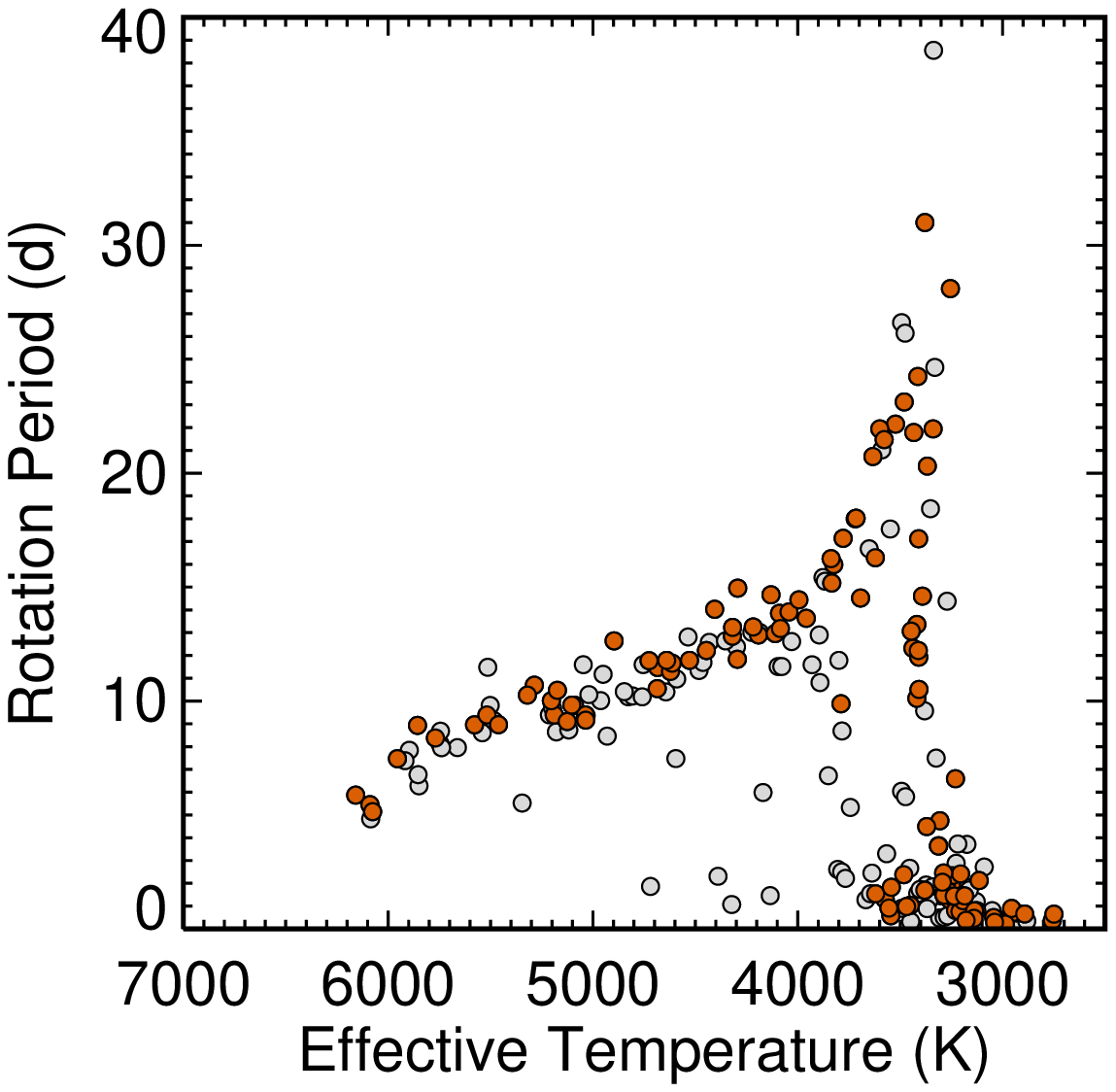}
\\ 
\includegraphics[trim=1.2cm 0cm 0.3cm 0cm, clip=True,  width=3.5in]{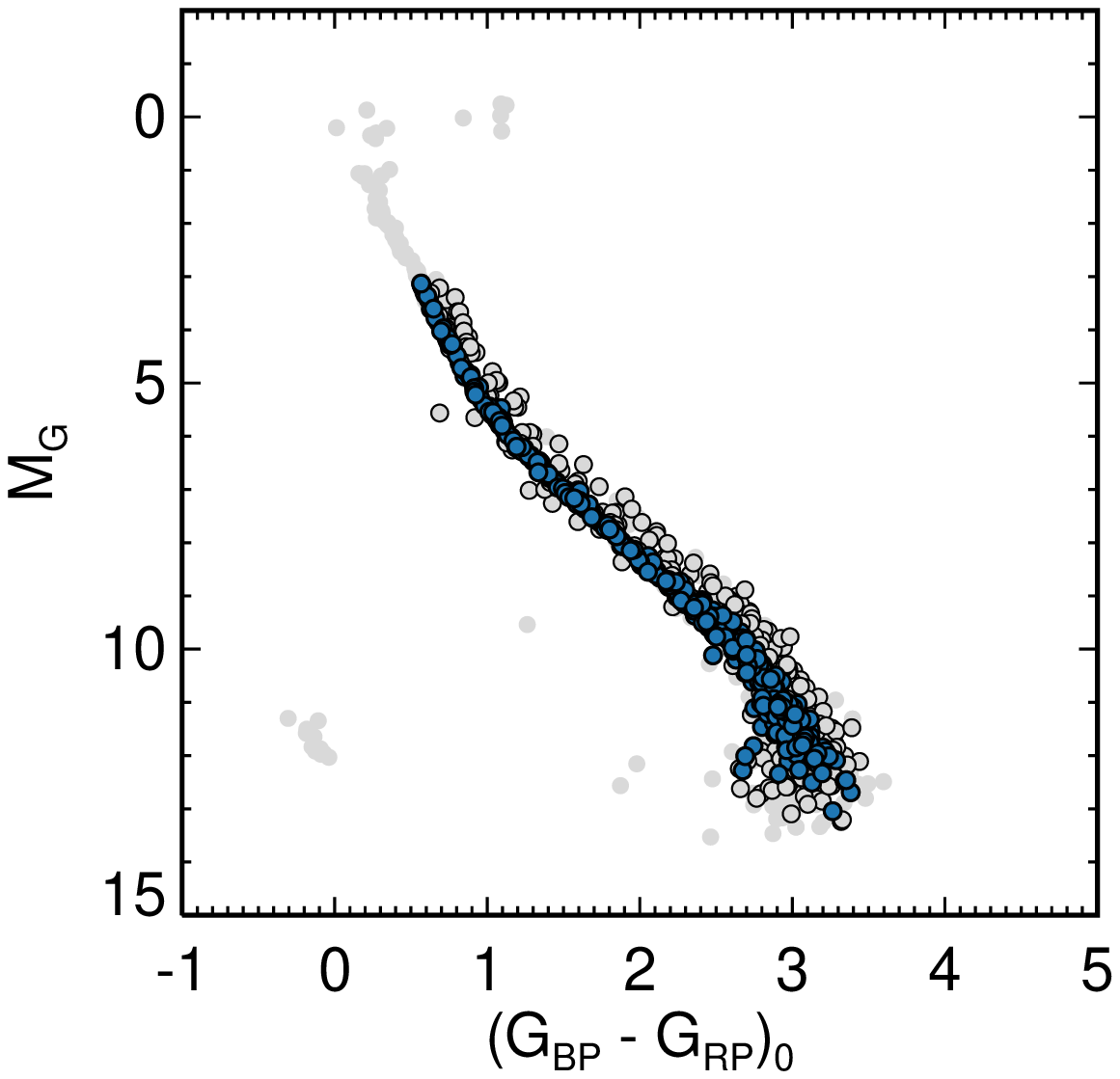}
\includegraphics[trim=1.2cm 0cm 0.3cm 0cm, clip=True,  width=3.5in]{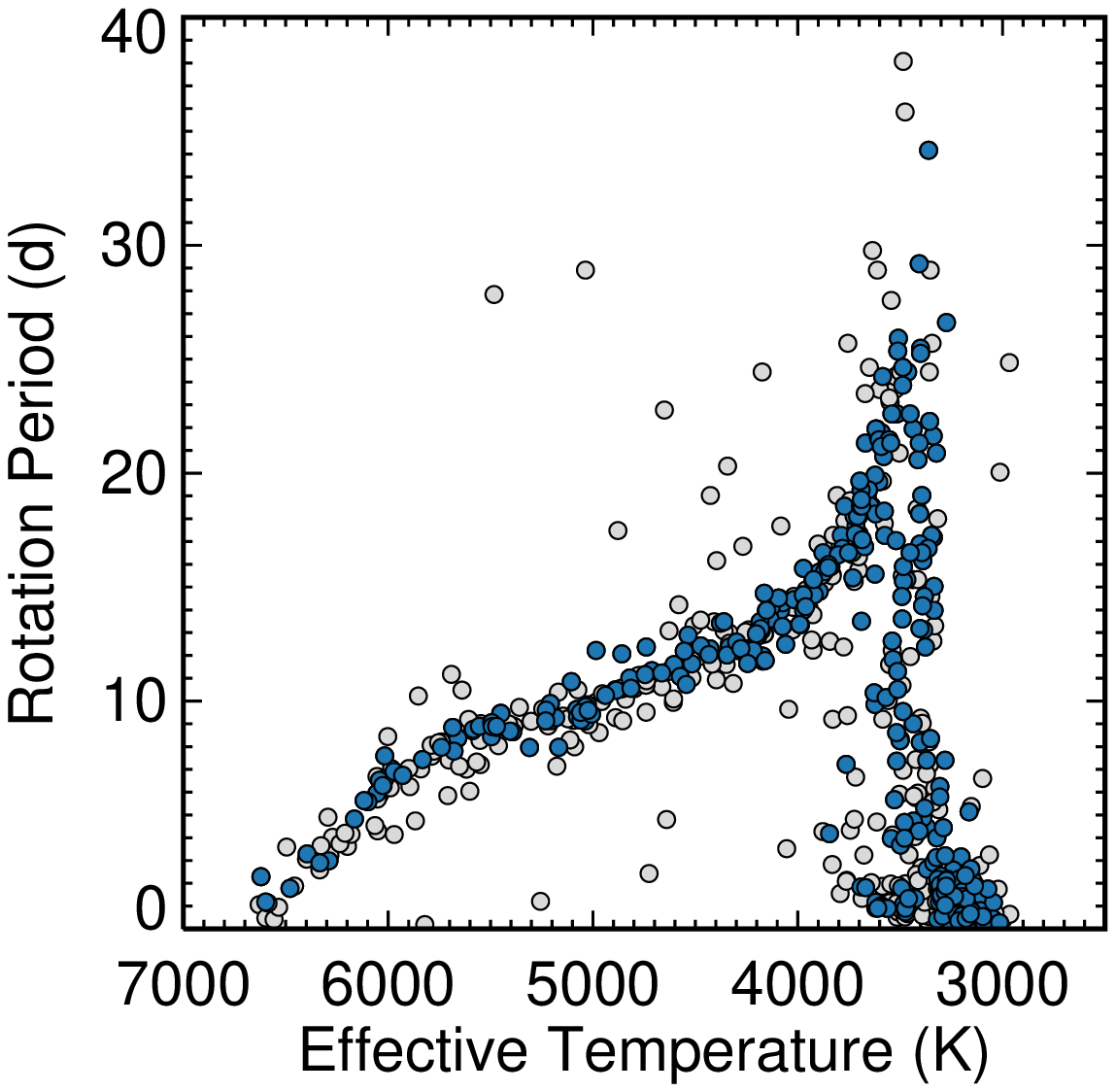}
  \caption{CMDs (\textit{left column}) and $P_{rot}$ distributions (\textit{right column}) for the Hyades (\textit{top row}) and Praesepe (\textit{bottom row}).
  \textit{Top left---}\citet{DR2HRD} Hyades members are plotted as gray points, stars with measured $P_{rot}$ are gray points outlined with black circles, and the subset we identify as single-star members are shaded orange.
  \textit{Top right---}$P_{rot}$ for the Hyades are plotted against \teff, which we calculated from \textit{Gaia} DR2 photometry using empirical color--temperature relations described in Section~\ref{temp}.
  Filtering out rotators that are known spectroscopic binaries,
  or have multiple periods detected in \textit{K2} light curves,
  or are astrometric or photometric non-single members from \textit{Gaia} DR2 removes all rapid outliers from the diagram, revealing a cleanly converged
  slow sequence down to $\teff \approx 3500$~K.
  \textit{Bottom left---}\citet{DR2HRD} Praesepe members are plotted as gray points,
  stars with $P_{rot}$ are gray points outlined with black circles,
  and the subset we consider to be single-star members are shaded blue.
  \textit{Bottom right---}
  Applying similar filtering as described for the Hyades
  removes all slow outliers in Praesepe's $P_{rot}$ distribution, and all rapid outliers down to $\teff \approx 4000$~K. The slow sequence appears converged
  down to $\teff \approx$ 3600-3800~K,
  depending on the still uncertain multiplicity of a few rapid stars in that
  range.
    \label{f:cluster_trim}}
\end{center}\end{figure*}

\subsection{Defining Single-Star Sequences}\label{single}
A comparison of Figure~\ref{fig:periodmass_k2} and figure~7 in \cite{douglas2017} shows that the color--$P_{rot}$ distributions for the Hyades and Praesepe appear qualitatively similar to each other. Most stars follow a common slow-rotator sequence from the late-F stars down to early M, followed by a sharp transition from slow to rapid near the fully convective boundary at $\approx$M4. However, many stars are outliers and appear to be rotating more rapidly or slowly than the slow-rotating sequence.

Where possible, it is important to reject outliers following membership and multiplicity criteria, instead of removing them based on their position in color--period space.
The primary reason is that we wish to show that $\approx$700-Myr-old stars follow a single-valued color--$P_{rot}$ relation from mid-F down to early M,
and that any rapid stars in this mass range are rapid for a reason unrelated to single-star angular-momentum evolution (e.g., because they are binaries, blends, or interlopers, or have poor data). 
Since the Hyades and Praesepe samples of rotators are large, we can apply strict physical (e.g., based on positions, kinematics, or luminosity excesses) and data-quality criteria (e.g., poor astrometric solutions, blended light curves resulting in multiple period detections) 
to select stars with \textit{Kepler} and \textit{Gaia} data consistent with single-star membership without over-depleting the color--period plane at any color. We describe our selection criteria below; each criterion is applied independently and the outputs combined to create our final list of single members. The results are summarized in Figure~\ref{f:cluster_trim}, and Tables~\ref{tab:k2} and \ref{tab:prae} include flags indicating which tests were passed by each star 

\subsubsection{Kinematics}
For the Hyades,
we select candidate single
stars first by rejecting confirmed binaries identified in the literature,
and then by considering the Galactic $UVW$ space velocities for stars with six-parameter positions and kinematics from \textit{Gaia} DR2.
We calculate the cluster median $UVW$ velocities from the Hyades membership list in \citet{DR2HRD};\footnote{For the Hyades,  $(U,V,W) = (+42.3, -19.2, -1.2)$~\kms.}
next, we compute the absolute velocity deviation,
$\Delta v$, for the 101 rotators in our sample 
with six-parameter positions and kinematics  
by subtracting off the cluster median values for each $UVW$ component and then adding the residuals in quadrature.
The Hyades's internal velocity dispersion is estimated to
be only 0.3 \kms\
\citep{Gunn1988, perryman1998},
which is comparable to the DR2 radial velocity (RV) error.
We adopt a more conservative threshold for identifying non-single members of $\Delta v$~$>$~2~\kms,
which eliminates 26 stars.
We also consider stars with DR2 RV errors $\sigma_{RV}$~$>$~2~\kms\
to be non-single members,
which cuts an additional four stars, so that in the end we have 71 single-star rotators in our sample. 

For Praesepe, we first remove the 43 binaries confirmed in the literature. Then, we filter non-single-member stars using proper motions separately from RVs. This is possible because Praesepe is more distant than the Hyades and useful because 719 of 743 rotators have DR2 proper motions, whereas only 185 have DR2 RVs.
The distribution of proper motions for our rotator sample can be approximately described by a Gaussian with $\sigma = 1.25$ \mas (the median proper motion error is 0.2 \mas). We set our threshold at twice this value and reject stars with absolute proper motion deviations larger than this 2.5~\mas.\footnote{For reference, a 0.5 \kms\ velocity dispersion at the distance of Praesepe (186~pc) translates to $\approx$0.57 \mas.} 
This eliminates 146 of 719 stars with DR2 proper motions. Separately, we
reject 48 stars with $\Delta$RV~$>$~2~\kms\ from the cluster
median value quoted by \citet{DR2HRD},\footnote{For Praesepe, $\mu_{\alpha} \cos \delta, \mu_{\delta} = \{ -36.047, -12.917 \}$ \mas.}
and 46 stars with $\sigma_{RV}$~$>$~2~\kms. In total, we reject 196 unique non-single members, and retain 523 single-star rotators.

\begin{deluxetable*}{llrclll}[t]
\tablewidth{0pt}
\tabletypesize{\scriptsize}
\tablecaption{Praesepe stars with measured $P_{rot}$ \label{tab:prae}}
\tablehead{\colhead{2MASS} & \colhead{EPIC} & \colhead{$P_{rot}$ (d)} &
\colhead{P\tablenotemark{a}} & \colhead{$T_{eff}$} & \colhead{DR2Name\tablenotemark{b}} & \colhead{SingleFlag}
}
\startdata
\nodata & 212004731 & 3.96 & 2 & 6196.83 & 661438260802777984 & NYYNN \\
\nodata & 211930461 & 14.59 & 2 & 4019.60 & 661211147230556032 & YYYNN \\
08410747+2154567 & 212094548 & 6.60 & 2 & 3097.71 & 665178391340402944 & NYY-Y \\
08395507+2003542 & 211988287 & 3.29 & 2 & 6395.48 & 664327433763175040 & YYYYY \\
08400063+1948235 & 211971871 & 2.99 & K & 6289.11 & 661311752544248960 & YYY-Y \\
08400130+2008082 & 211992776 & 1.18 & 2 & 6597.70 & 664328915529294976 & YYY-Y \\
08402232+2006244 & 211990908 & 2.59 & K & 6333.77 & 661419259867454976 & YYNYN \\
08401763+1947152 & 211970750 & 6.69 & K & 6054.54 & 661310790468509952 & -NY-Y
\enddata
\tablenotetext{a}{Flag for the $P_{rot}$ source selected. ``S'':\citet{scholz2007,scholz2011}, ``P'':\citet{agueros11} (PTF), ``D'':\citet{delorme2011} (SWASP), ``K'':\citet{kovacs2014} (HATnet), ``2'':\cite{douglas2017} (\textit{K2} Campaign 5), and ``16'': this work (\textit{K2} Campaign 16; see Section~\ref{prot_correct}) }
\tablenotetext{b}{ Flags for selecting single stars in Section~\ref{single}. Entries correspond to astrometry, photometry, \textit{K2} multiperiodic, RV, and confirmed binary selection: ``Y'' indicates the star passes a given test, ``N'' indicates failure, and ``-'' indicates that we lack the data to perform a particular test.  We only retain stars flagged ``YYYYY'' or ``YYY-Y''. }
\end{deluxetable*}








\subsubsection{Photometry}\label{phot_selection}
We use the \citet{DR2HRD} Hyades catalog to generate a fiducial cluster CMD, and then iteratively fit the resulting main-sequence with a cubic basis-spline.
We then generate a new CMD using our full rotator list, and determine each star's deviation from the fiducial main-sequence.

We fit two CMDs: absolute $G$ magnitude, $M_G$,
versus both \gbr\ and ($G - G_{\rm RP}$).
We analyze ($G - G_{\rm RP}$) to
account for the larger uncertainty in
$G_{\rm BP}$ for redder/fainter stars.
We then calculate the photometric deviation from these empirical
main-sequences for our rotator sample,
$d_{cmd} = |M_{G,observed} - M_{G,predicted}|$,
and label all stars that are consistent with at least one
of the empirical isochrones as
photometric single-member stars.
We set a threshold of $d_{cmd} < 0.375$~mag for all stars,
which is half of the offset for an equal-mass binary \citep[e.g.,][]{hodgkin1999}.
We find that 176 of 222 Hyads with DR2 photometry are consistent with being single-star members.

For Praesepe, we adjust the fiducial Hyades CMD fit
according to its interstellar reddening/extinction that we
derive in Appendix~\ref{reddening} ($A_V = 0.035$)
and the distance modulus
implied by inverting the cluster parallax
\citep[$\varpi = 5.371$ mas;][]{DR2HRD}.
We find 525 of the 741  stars with DR2 photometry
are consistent with being single-star members.

\subsubsection{Astrometric data quality}
The \textit{Gaia} DR2 astrometric solution for each star assumes it is a single
point source. Objects that are inconsistent with this assumption
can have excess astrometric noise ($\epsilon_i$), and
we remove those with $\epsilon_i > 1$ and $G < 19$~mag from our samples.
This includes 40 stars in the Hyades
and 48 stars in Praesepe.
Most were already filtered by our kinematic and
photometric selection criteria.

\subsubsection{$P_{rot}$ quality and corrections}\label{prot_correct}
For rotators with \textit{K2} light curves, we remove those for which we detect multiple periods, which, again, we interpret
as either physically unassociated blends or cluster binaries (see Sections~\ref{binaries} and \ref{validation}). 

For Praesepe, an additional step is required: several periods in the literature need to be corrected.
In \citet{douglas2017}, we assembled literature periods and our own \textit{K2} periods, and then recommended which source to use for each star.
We recommended \citet{delorme2011} for EPIC 211995288,
and
\citet{scholz2007} for EPIC 211970147 \citep[K2-102;][]{mann2017}. 
But after re-inspecting the \textit{K2} light curves, it is clear that our \textit{K2} periods are accurate and the literature values are half-period harmonics.

The Campaign 5 light curves for EPIC 211890774  
and EPIC 211822797 \citep[K2-103;][]{mann2017}
both show weak asymmetries in the depths of alternating minima, which we confirm with their Campaign 16 light curves. This indicates that the \citet{douglas2017} measurements for these two stars are half-period harmonics, caused by presumably by nearly symmetric spot patterns on opposite-facing hemispheres. We therefore double the old $P_{rot}$ for these stars. 

Finally, EPIC 211950227 
was originally
given a period of 13.15 d \citep{delorme2011}.
However, the Campaign 16 light curve shows that the dominant modulation signal
has a period of $P_{rot} = 1.76$ d.
We see no $\approx$13~d signature in its Campaign~16 light curve and conclude that the {\it K2}-derived $P_{rot}$ is the correct one.

\subsection{Resulting CMDs and \teff--$P_{rot}$ Distributions for the Hyades and Praesepe}
The resulting CMDs for the two clusters are shown in the left column of Figure~\ref{f:cluster_trim}, with their $\teff$--$P_{rot}$ distributions in the right column. Applying the cuts described above yields a nearly clean $P_{rot}$ distribution for both clusters. Overall, 118 Hyades rotators out of 232 satisfy our single-star-membership criteria.
When examining the cluster's $P_{rot}$ distribution, we find no rapid outliers relative to the cleaned, slow-rotating  sequence for $M_\star \gtrsim 0.57$~\msun\ ($\teff \gtrsim 3789$~K), and only three moderately faster rotators for $M_\star \gtrsim 0.5$~\msun\ ($\teff \gtrsim 3620$~K). The transition to completely rapid rotation in the Hyades 
occurs at $M_\star \approx 0.35$~\msun\ ($\teff \approx 3420$~K, M3), which is slightly warmer than the \teff--radius discontinuity at \teff~=~3200--3340~K identified by \citet{Rabus2019}.

For Praesepe, we find that 496 of the 743 rotators
are consistent with being single-star members.
None of these stars appears significantly more rapid than the slow converged sequence for $\teff \gtrsim 3845$~K ($M_\star \gtrsim 0.6$~\msun, M0). Of the 43 single members on our list with $3600 < \teff < 3850$~K, 10, or 23\%, are rapidly rotating outliers that have $P_{rot}$ faster than the slow sequence by at least 3~d. The transition to all rapid rotators happens around $M_\star \approx 0.4$~\msun\ ($\teff \approx$~3500~K), but is not as well defined as in the Hyades.

Finally, Pr0211 (EPIC 211936827, \textit{Gaia} DR2 \\661222279785743616) hosts a hot Jupiter \citep[$M_p \sin i = 1.844$~\mjup, $\porb = 2.15$~d;][]{quinn2012}.
We find that Pr0211 rotates 1.4~d (15\%) faster than expected, in agreement with \citet{kovacs2014}

\subsection{A Precise Differential Gyrochronology Age for the Hyades}\label{diff}
We now turn to the question of whether Praesepe and the Hyades are truly coeval. 
We search the literature and tabulate recent isochrone ages for the two clusters derived using a variety of photometry, constraints, models, and methods (see Table~\ref{table:ages}). From these, we calculate an age for the Hyades of 728$\pm$71 Myr (median and 1$\sigma$ of thirteen values), and for Praesepe of 670$\pm$67~Myr (median and 1$\sigma$ of eleven values). 
Since this suggests that Praesepe is the younger of the two clusters, we then calibrate an empirical gyrochronology model by fitting the Praesepe  $\teff$--$P_{rot}$ sequence, and then tune the age-dependence with the Sun.
Finally, we compare the $\teff$--$P_{rot}$ sequences of the Hyades and Praesepe, and derive a precise, differential age according to our empirical model.

We summarize our assumed values for the Sun here.
We take the Sun's $P_{rot}=26.09$~d, measured from periodic modulations in the Mount Wilson \caiihk\ S-index by \citet{donahue1996}.
We take its age to be 4567$\pm$1$\pm$5~Myr \citep{ageofsun}.
Based on observations of solar twins derived from the updated Spectroscopic Properties of Cool Stars \citep[SPOCS;][]{Brewer2016} catalog, we derive a solar color of
($G_{\rm BP} - G_{\rm RP}$)$_\odot$~=~0.817~mag, consistent with the value of
($G_{\rm BP} - G_{\rm RP}$)$_\odot$~=~0.82 estimated by \citet{Casagrande2018}.
A more detailed discussion of 
our derivation of 
this color can be found in Appendix~\ref{sun}.

Our analysis also makes the following assumptions:
\begin{enumerate}
    \item The Sun has slowed down continuously since it was 670~Myr old (our adopted age of Praesepe). According to \citet{vansaders2016}, magnetic braking efficiency plummets at a critical Rossby number (the ratio of $P_{rot}$ to convective turnover time) of $R_{\rm crit} = 2$, approximately the current solar value. We assume that the Sun has not yet reached this threshold and that it has therefore spun down continuously with a single-valued time dependence.
    \item The difference in metallicity between the Sun and Praesepe does not appreciably affect spin-down and that
    comparing equal-color stars is valid, even though a solar-mass star in Praesepe is cooler than the Sun's current temperature.\footnote{Stars do not spin down through $\teff$--$P_{rot}$ space along perfectly vertical lines, since they warm as they age. Differences in metallicity will also modify moments of inertia, convective turnover times, and other physical ingredients that are critical to understanding angular-momentum evolution.
    Theoretical models are the appropriate way of accounting for metallicity and stellar-evolution effects \citep[e.g.,][]{vansaders2013}, but we presently lack sufficient coeval benchmarks with different metallicities to validate their predictions. Also, all available models fail to represent the cluster sample, aside from the most Sun-like G~dwarfs \citep[e.g.,][Curtis et al.~submitted, and this work]{Agueros2018}. 
    Since our primary goals are to test if the Hyades and Praesepe are truly coeval and to measure a differential age, any systematic inaccuracies in the model will propagate to both cluster ages equally.}
\end{enumerate}

\begin{figure*}\begin{center}
\includegraphics[trim=0.6cm 0cm 0.3cm 0cm, clip=True,  width=3.5in]{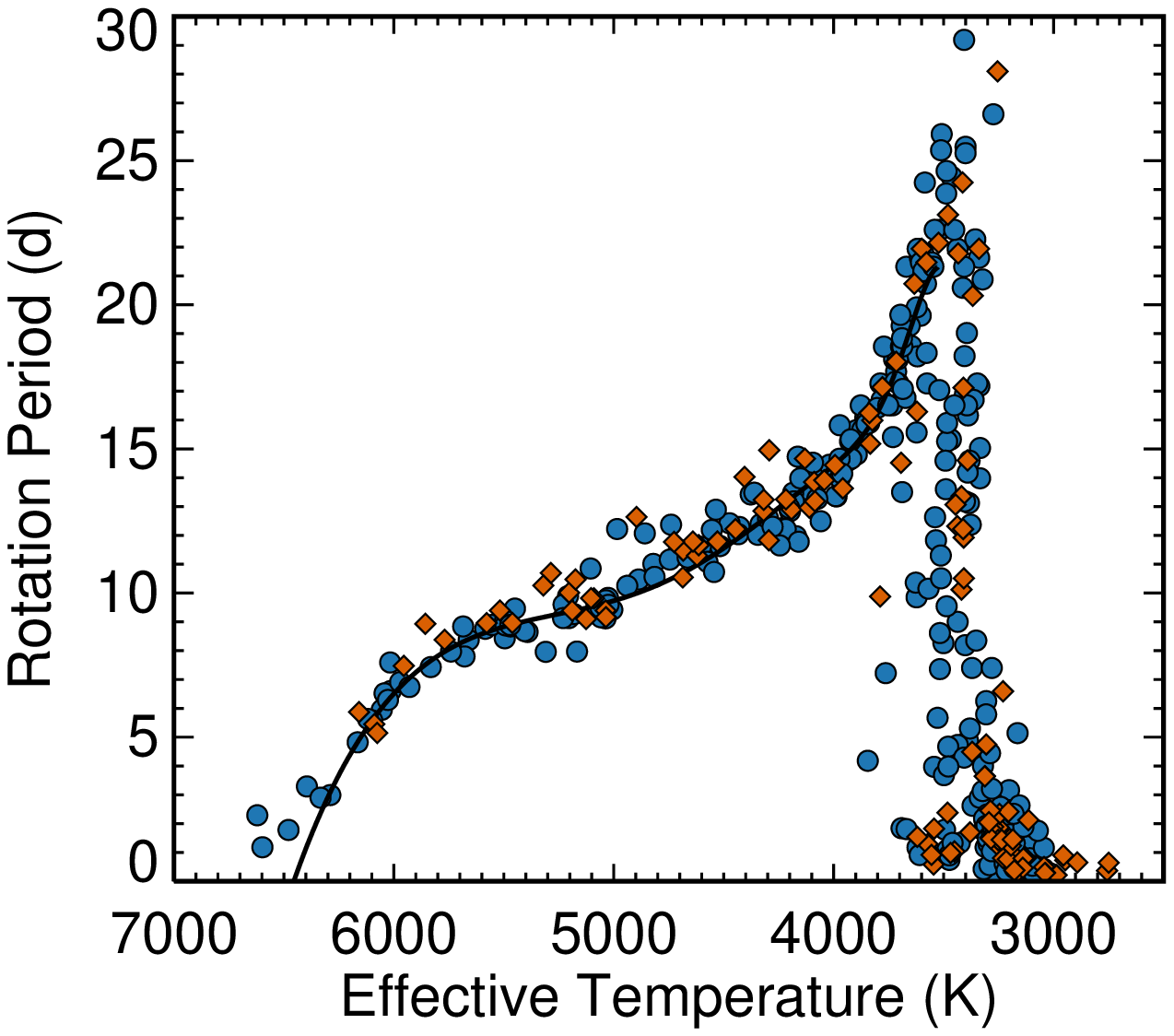}
\includegraphics[trim=0.6cm 0cm 0.3cm 0cm, clip=True,  width=3.5in]{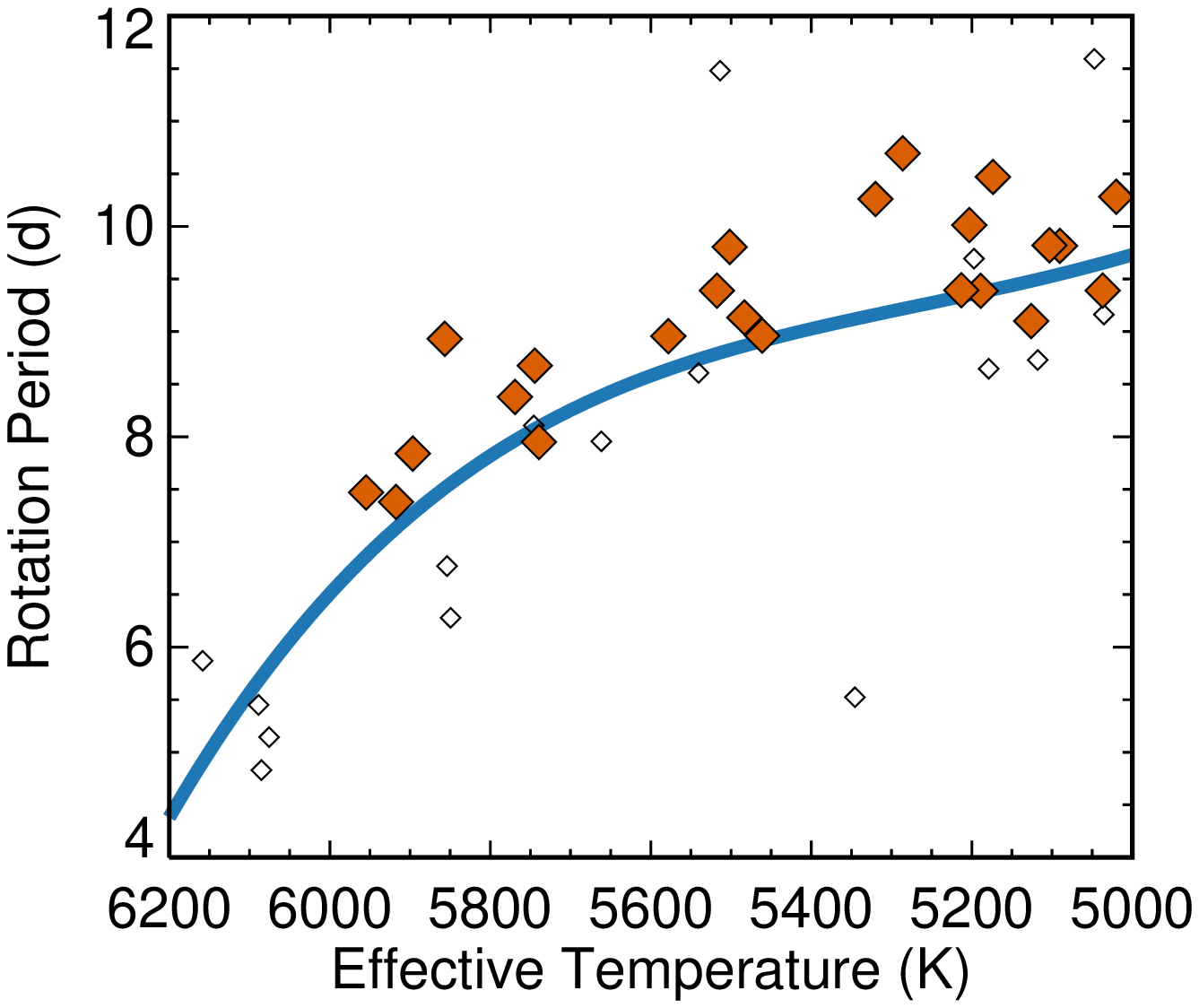}
\\
\includegraphics[trim=0.6cm 0cm 0.3cm 0cm, clip=True,  width=3.5in]{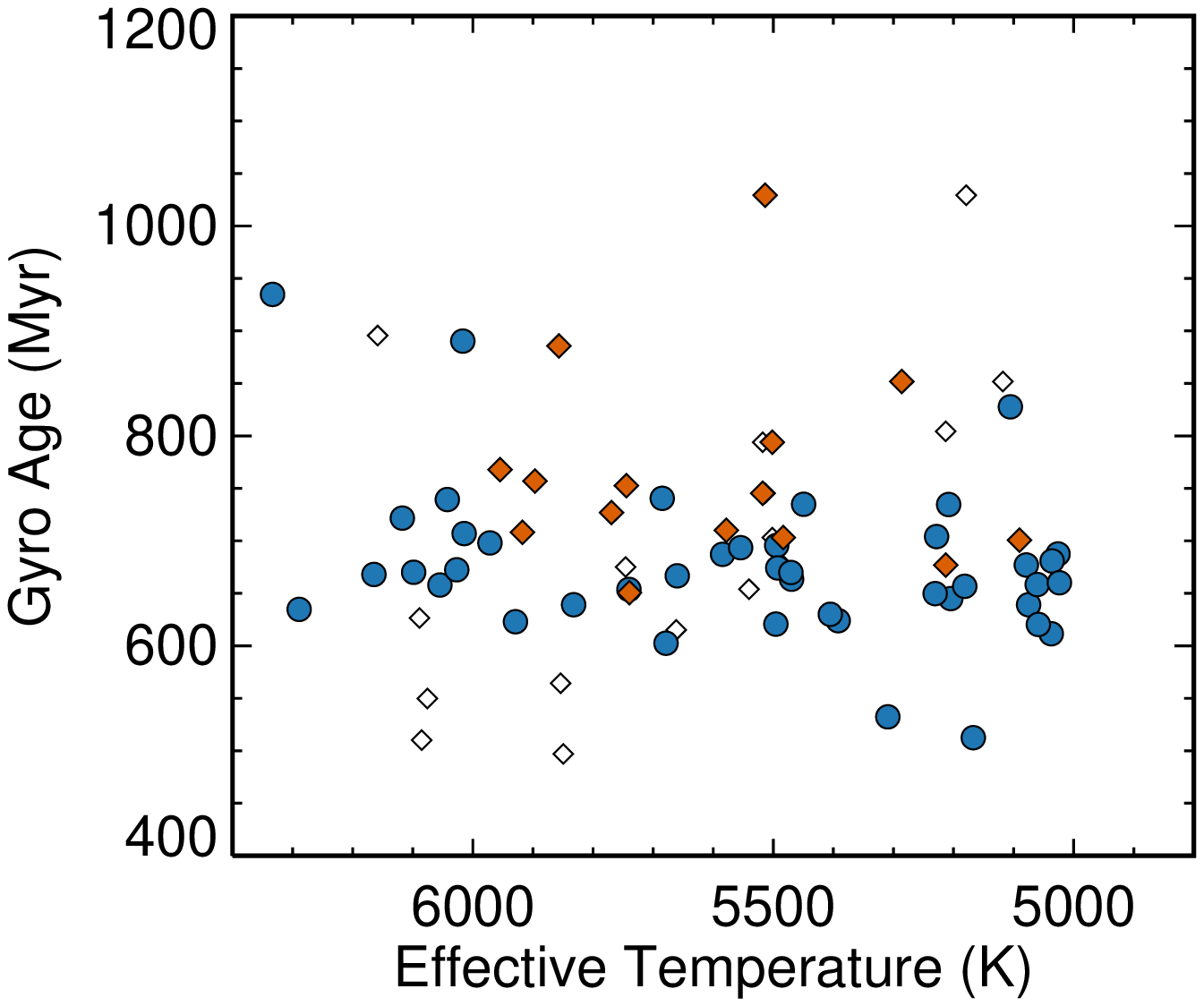}
\includegraphics[trim=0.6cm 0cm 0.3cm 0cm, clip=True,  width=3.5in]{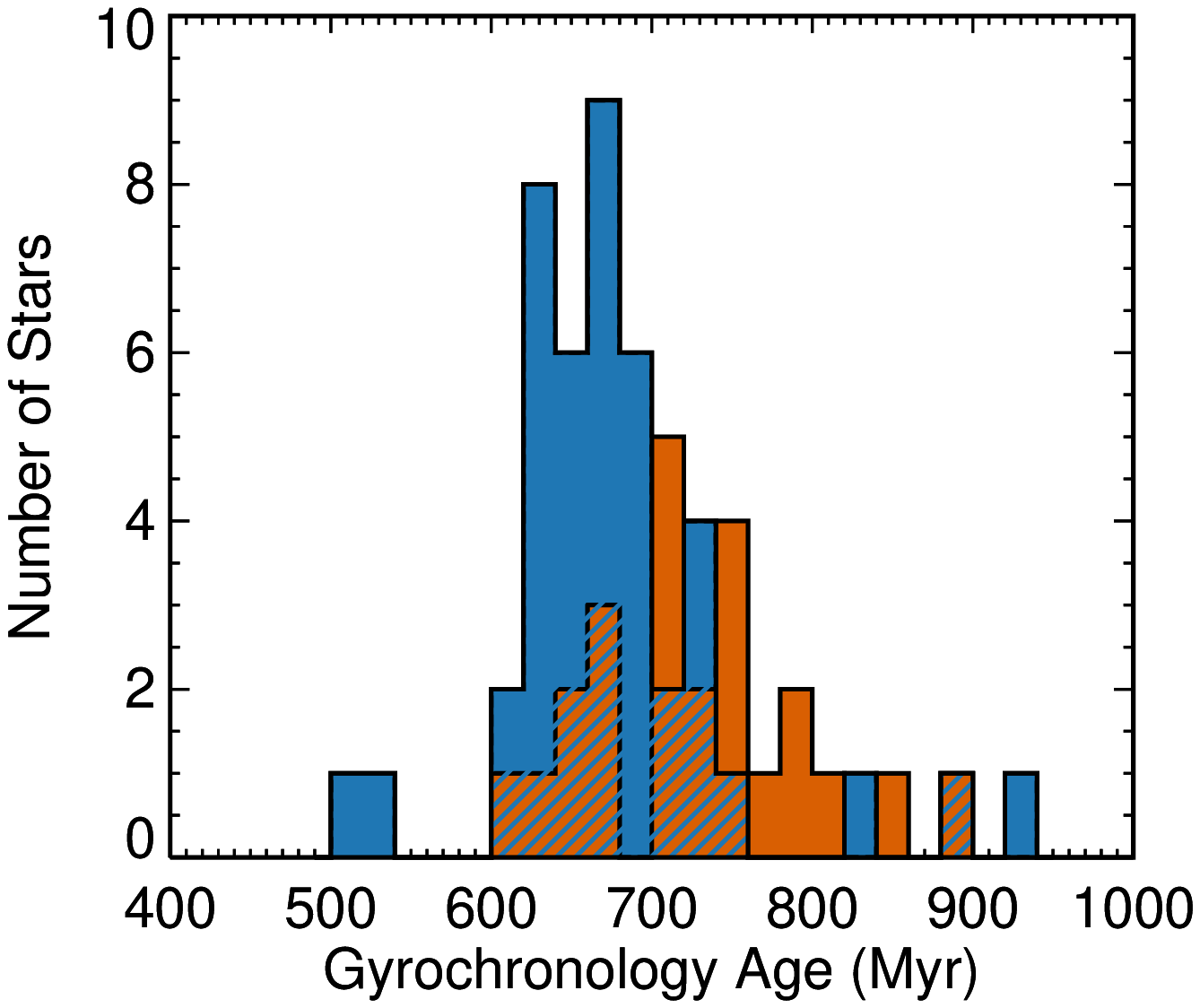}
  \caption{\textit{Top left---}The $\teff$--$P_{rot}$ distributions for likely single stars in
  Praesepe (blue circles) and the Hyades (orange diamonds).
  The black line is a polynomial fit to the Praesepe stars
  with $0.65 < (G_{BP} - G_{RP})_0 < 2.4$.
  The distributions for the two clusters appear roughly consistent.
  \textit{Top right---}Rotation data for single Hyads (orange diamonds) used in the gyrochronology age calculation ($5000 < \teff < 6000$~K). Non-single Hyads (open diamonds) are shown for comparison.
  Single Hyads rotate systematically more slowly than the
  Praesepe polynomial model (blue line).
  \textit{Bottom left---}Gyrochronology ages for the single Praesepe (blue circles above) and Hyades (orange diamonds above) samples
  over the range in \teff\ where gyrochronology should be viable at this age
  \citep{Agueros2018}.
  These Hyads are systematically older than the Praesepe stars (adopted age of 670 Myr) by $\approx$57 Myr.
  \textit{Bottom right---}A histogram of the data in the scatter plot in the previous panel.
 The age distributions overlap, but the Hyades sample is systematically older by 57 Myr.
 Having large samples of stars helps mitigate the uncertainties for individual stars that caused the apparent age spreads (whether due to $P_{rot}$ measurement uncertainties or intrinsic spread) 
 for age-dating coeval populations.
    \label{f:gage}}
\end{center}\end{figure*}

We fit a sixth order polynomial to Praesepe's cleaned and dereddened DR2 color--period sequence for stars with ($\gbr$)~$< 2.4$
($\teff \approx 3500$~K, $M_\star \approx 0.42$~\msun, M2V).
This color limit stops our model before the sharp drop to rapid rotation around the fully convective boundary.
The sixth order polynomial is necessary as lower-order polynomials fail to accurately track the rapid change in $P_{rot}$ from the F to G dwarfs. 


The Praesepe fit predicts a period at the solar color
of $P_{rot} = 8.09$$\pm$$0.25$ d.
We calculate this value using a \teff--$P_{rot}$ diagram de-reddened by our $A_V=0.035$ value, while the uncertainty comes from assuming either $A_V = 0$ (no reddening) or $A_V = 0.084$ \citep{taylor2006}.
We use the age for Praesepe derived from the literature of 670~Myr, and calculate that the 
braking index $n = 0.619$.

We now apply our new gyrochronology formula
to the cleaned stars in the Hyades
with $0.7 < $($G_{BP} - G_{RP}$)$_0 < 1.1$, 
where gyrochronology should be viable at this age \citep[][Curtis et al.~in prep.]{Agueros2018}.
If Praesepe is 670 Myr old and
its $A_V = 0.035$, and if it is chemically identical to the Hyades, then the Hyades is 57 Myr older. We find the Hyades age to be 727$\pm$75 Myr (median and 1$\sigma$), based on 25 cluster members.
(For 49 analogous Praesepe stars, 1$\sigma$ = 69~Myr.)
Recall that we calculate an isochrone age difference of 58 Myr by
computing the difference between the median of various isochronal ages for each cluster;
this is essentially identical to our differential gyrochronology result.

Figure \ref{f:gage} shows the
\teff--$P_{rot}$ diagram for the cleaned
Praesepe and Hyades samples,
and their corresponding gyrochronology ages
using our recalibrated formula. 
Derived ages for individual stars are given in Table~\ref{tab:ages}.

\section{Discussion}\label{res}

New $P_{rot}$ measurements from \textit{K2} and precise \textit{Gaia} data have enabled us to compare the rotation distributions in Praesepe and the Hyades in detail.
Whereas in previous work we assumed that the clusters have overlapping $P_{rot}$ sequences, we now find that is not the case for solar-type stars.
Overall, we find that Hyades FG stars rotate more slowly than their Praesepe counterparts, corresponding to a differential gyrochronological age of 57~Myr.
This difference is consistent with the $47\pm17$~Myr difference between the clusters found by \citet{delorme2011}, who used a linear fit to the $P_{rot}$ vs. ($J-K_s$) relation in the Hyades and Praesepe.
The 57~Myr age difference suggests that the two clusters should be separated when considering the evolution or effects of stellar rotation in solar-type stars and when accuracy below the 10\% level is required.

\begin{figure}\begin{center}
\includegraphics[trim=0.6cm 0cm 0.3cm 0cm, clip=True,  width=3.5in]{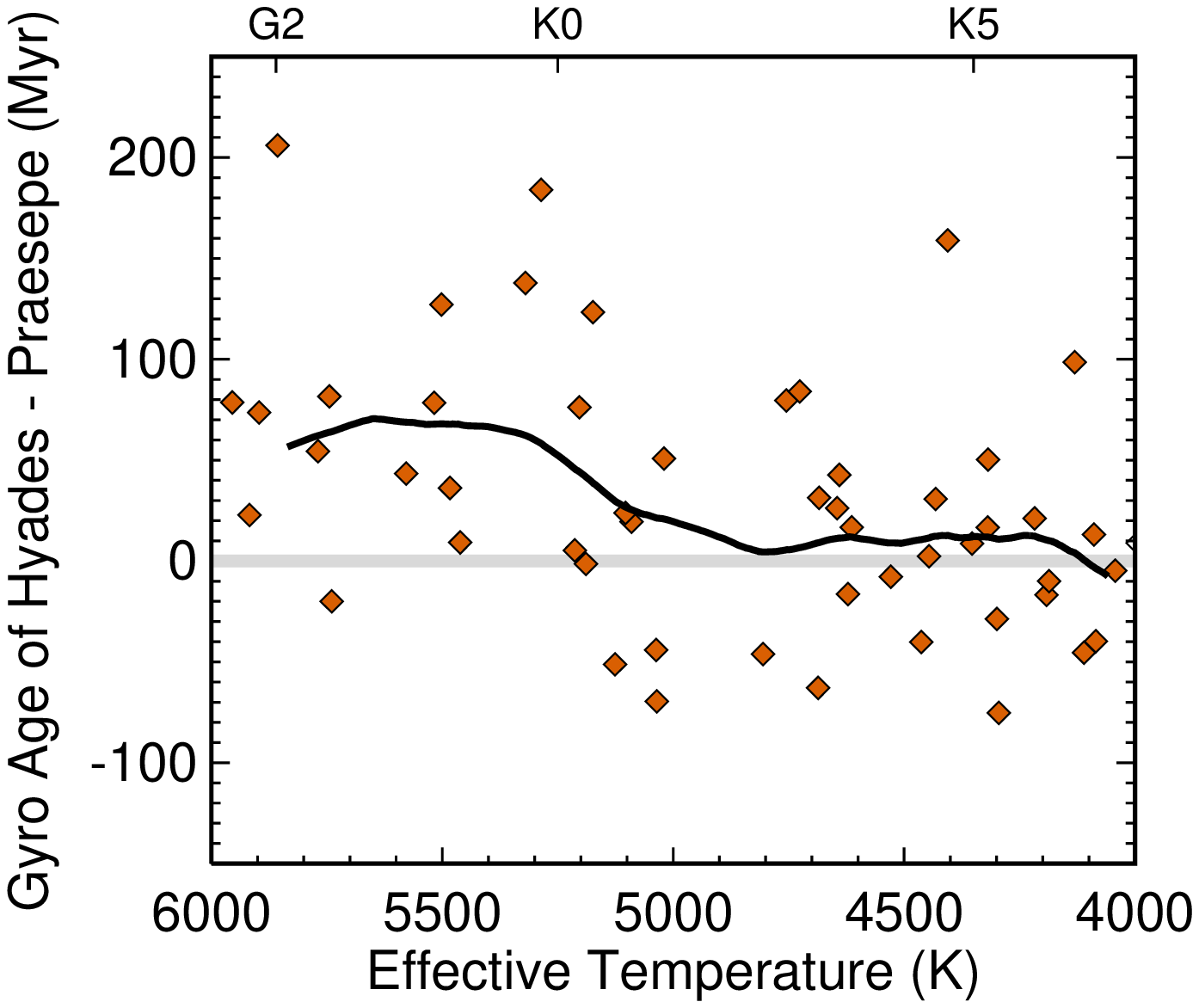}
  \caption{The differential gyrochronology ages for individual Hyads (orange diamonds) compared to our fiducial Praesepe model (grey horizontal line).
  We also show the LOWESS regression for the age difference, showing that the difference in gyrochronology ages is strongest for G stars, and decreases between $5250 > \teff > 4900$~K until the cooler Hyads appear coeval with their cousins in Praesepe.
    \label{lowess}}
\end{center}\end{figure}

Interestingly, the age discrepancy between the two clusters is largest for $\teff > 5200$~K and decreases as we move to cooler stars.
We fit the gyrochronology ages of Hyades stars with locally weighted scatterplot smoothing (LOWESS) as a function of \teff, and compare it to the fiducial Praesepe model (Figure~\ref{lowess}).
Between $5250 > \teff > 4900$~K, the differential gyro ages decrease, so that cooler Hyads converge with the Praesepe sequence.
The late K and early M dwarfs do not brake appreciably from the age of Praesepe to that of the Hyades. 
This contradicts the common assumption 
that braking timescales increase as mass decreases.
Our work therefore adds to prior evidence that low-mass stars follow a different, more complex braking timeline than their solar-type counterparts.

Several other authors have reached similar conclusions.
\citeauthor{meibom2009} 
compared
M35 \citep[$\approx$150~Myr;][]{meibom2009},
M34 \citep[$\approx$220~Myr;][]{meibom2011-1},
and NGC 6811 \citep[$\approx$1~Gyr;][]{meibom2011} to the Hyades, and find that K dwarfs must spin down less efficiently than FG stars. 
\citet{cargile2014} found the same result by comparing Blanco~1 and the Pleiades (both $\approx125$~Myr) to M37 ($\approx550$~Myr), the Hyades, and NGC~6811.
Similarly, \citet{Agueros2018} found evidence for stalling from the age of Praesepe to that of NGC 752 ($\approx$1.3 Gyr) for K and early M stars. Finally, Curtis et al.\ (submitted) re-examined NGC~6811 by searching \textit{Gaia} DR2 for additional members with \textit{Kepler} light curves, thereby significantly expanding the size of that cluster's rotator sample
and extending it down in mass from $M_\star \approx 0.8$~\msun\ to
$\approx$0.6~\msun. 
Surprisingly, these authors found that NGC~6811's slow rotator sequence converges with that of the Hyades and Praesepe at redder colors, indicating that these stars effectively do not spin down at all over a time span of several 100~Myr.

We therefore provide concrete evidence that K stars spin down at a variable rate, as opposed to existing empirical  models which show them spinning down continuously from the time they reach the main sequence. 
This stalling is apparent even over $\sim$50~Myr timescales.
Previous empirical work has assumed a fixed functional form for the dependence of $P_{rot}$ on mass or ($B-V$) at all ages.
For example, \citet{delorme2011} fit a line to the $P_{rot}$ vs. ($J-K_s$) distributions in clusters, 
and 
\citet{barnes2003,barnes2007} fit 
other analytic functions.  
These efforts assumed that it was possible to decouple the mass and age dependencies, but our results demonstrate that rotation evolves at different rates for stars of different masses.

\citet{barnes2010} presented the only empirical gyrochronology relation that allowed more complicated mass-dependent evolution by including a dependence on the convective turnover time $\tau$, instead of color.
That model accurately described the 
$M_\star > 0.85$~\msun\ stars in the 2.5~Gyr NGC~6819 cluster \citep{meibom2015}. 
However, it actually predicted that K dwarfs 
spin down more rapidly than G dwarfs, 
instead of more gradually as indicated by the 
open cluster data.
\citet{mamajek2008}, \citet{meibom2009}, and, more recently, \citet{angus2015} simply re-calibrated the model presented by \citet{barnes2003,barnes2007}, without considering more complex mass-dependent rotational evolution.

One probable reason that empirical models have not included a mass dependence
is the paucity of $\gapprox$1-Gyr-old benchmarks for K and M dwarf rotators.
$P_{rot}$ have been published for solar-type members of NGC~6819 and M67, but not their lower-mass members.
We show that this dependence is present even over short timescales, but the field of gyrochronology requires additional benchmarks at older ages to properly calibrate braking timescales for stars of different masses.
Future work on NGC~6819 and Ruprecht 147, also $\approx$2.5 Gyr old, will provide further constraints on mass-dependent evolution at older ages.

For the time being, when examining effects at a single age,
we can consider the low-mass rotators in the Hyades and Praesepe as a single ensemble.
The low-mass rotators deserve additional consideration in future work, but this will first require comprehensive binary surveys of late K and early M dwarfs to disentangle evolutionary effects from multiplicity effects in these clusters.
Several authors have found tentative evidence that binaries rotate faster than single stars \citep[e.g.,][]{meibom2007,douglas2016,douglas2017}, which is one reason why we remove known binaries from our sample above.

The Hyades and Praesepe, however, have not been uniformly surveyed for binaries, particularly at the low-mass end.
In our \textit{K2} analysis, we identify candidate binaries via blends and multiple periods detected in a single light curve.
However, these candidates could be chance alignments or (when the two periods are very similar) a signal of latitudinal differential rotation.

NASA's ongoing \textit{Transiting Exoplanet Survey Satellite} mission \citep[\textit{TESS};][]{TESS} will also provide an excellent opportunity for expanding the $P_{rot}$ catalog for Hyades M dwarfs.
Many Hyades M dwarfs lie on the outskirts of the cluster \citep[with many more potentially found in unbound tidal tails;][]{roser2019},
far enough from the ecliptic to be observed by \textit{TESS}.
Although there will certainly be issues with systematics given the standard 27.4 d observing cadence,
we expect to measure $P_{rot}$ for $\approx$200 Hyads in the Southern Hemisphere alone (TESS Program G011197).
Many more Hyads, as well as members of another  approximately coeval Coma~Ber cluster \citep{collier2009}, will be observed by \textit{TESS} in the Northern Hemisphere.
Since one current challenge in comparing the Hyades and Praesepe is the much smaller Hyades $P_{rot}$ catalog, future \textit{TESS} measurements will be invaluable for differentiating the behavior of M dwarfs in these similarly aged clusters.

\section{Conclusions}\label{concl}

    We analyze \textit{K2} Campaign 13 light curves for 132 members of the Hyades open cluster.
    We measure $P_{rot}$ for 116 (88\%) of these stars, including 93 members with no prior $P_{rot}$ measurements, bringing the total number of Hyads with known $P_{rot}$ to 232.     
    As in our last two papers \citep{douglas2016,douglas2017}, we find  that ground-based $P_{rot}$ measurements are generally consistent with space-based measurements. The primary difference is that space-based observatories can observe a wide field of view nearly continuously while simultaneously reaching even faint members of nearby open clusters.

    We then use \textit{Gaia} DR2 data and literature binary information to define a clean sequence of single-star Hyads in color-magnitude space. We then apply this procedure to data for the Praesepe open cluster, which is generally thought to be coeval with the Hyades. As a result, we obtain two clean sequences of slowly rotating FGK stars in \teff--$P_{rot}$ space for both clusters.    

    There are far fewer known binaries among the M dwarfs in these two clusters. But our cuts also produce a nearly clean slow-rotator sequence for early M dwarfs, with only a few rapidly rotating members in this mass range in both clusters. These remaining rapid rotators highlight the need for additional binary surveys of M dwarfs in these clusters, especially Praesepe.

    We use these single-star sequences to derive a reddening value of $A_V = 0.035$$\pm$$0.011$~mag for Praesepe, assuming that the Hyades experiences no reddening. This value is intermediate between the oft-assumed $A_V = 0.0$ and the $A_V = 0.084$~mag derived by \citet{taylor2006} for Praesepe.
    We then derive a polynomial fit to the slow rotator sequence in Praesepe, as a function of dereddened \textit{Gaia} DR2 ($G_{\rm BP} - G_{\rm RP}$)$_0$ color. We use this fit as the basis for a new empirical model for gyrochronology, where we assume that stars begin on the Praesepe sequence at 670 Myr and their periods evolve as $P_{rot}\propto t^{n}$. By comparing the Praesepe sequence to the Sun, we derive a value of $n=0.619$.

    Finally, we compare the slow-rotator sequence in the Hyades to this model we have generated based on Praesepe. We find that, if we only consider the F and G stars, the Hyades is 57 Myr older than Praesepe. We also find, however, that the difference between the Hyades and Praesepe sequences decreases towards lower-mass stars, so that the K and early M dwarfs in the two clusters are indistinguishable. This provides further evidence for stalling in the rotational evolution of these stars, and highlights the need for more detailed analysis of spin-down over time for stars of different masses.

\acknowledgments

S.T.D.~acknowledges support provided by the NSF through grant AST-1701468.
J.L.C.~acknowledges support provided by the NSF through grant AST-1602662.
M.A.A.~acknowledges support provided by NASA through K2GO4 grant NNX17AF73G and by the NSF through grant AST-1255419.
We also thank the anonymous referee for comments which improved the manuscript.

We thank R.~Stefanik for sharing his Hyades binary detections with us. We also thank D.~Latham for contributing to those observations and for providing valuable advice.

This research has made use of NASA's Astrophysics Data System Bibliographic Services, the SIMBAD database \citep{simbad}, operated at CDS, Strasbourg, France, and the VizieR database of astronomical catalogs \citep{Ochsenbein2000}.

This paper includes data collected by the {\it K2} mission. Funding for the {\it K2} mission was provided by the NASA Science Mission directorate.


This work has made use of data from the European Space Agency (ESA) mission
{\it Gaia} (\url{https://www.cosmos.esa.int/gaia}), processed by the {\it Gaia}
Data Processing and Analysis Consortium (DPAC,
\url{https://www.cosmos.esa.int/web/gaia/dpac/consortium}). Funding for the DPAC
has been provided by national institutions, in particular the institutions
participating in the {\it Gaia} Multilateral Agreement.

This research has made use of the NASA/ IPAC Infrared Science Archive, which is operated by the Jet Propulsion Laboratory, California Institute of Technology, under contract with the National Aeronautics and Space Administration. The Two Micron All Sky Survey was a joint project of the University of Massachusetts and IPAC.

The Digitized Sky Survey was produced at the Space Telescope Science Institute under U.S. Government grant NAG W-2166. The images of these surveys are based on photographic data obtained using the Oschin Schmidt Telescope on Palomar Mountain and the UK Schmidt Telescope. The plates were processed into the present compressed digital form with the permission of these institutions.


\appendix

\section{Non-Zero Reddening in Praesepe}\label{reddening}

\begin{figure}\begin{center}
\includegraphics[trim=1.0cm 0cm 0.0cm 0cm, clip=True,  width=3.4in]{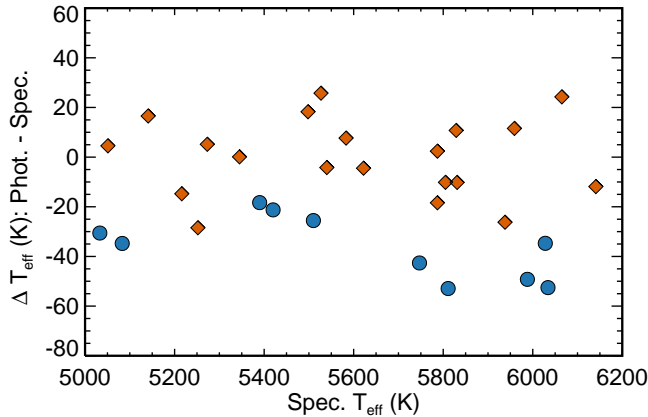}
  \caption{The difference between photometric ($T_{\rm eff, phot}$)
  and spectroscopic effective temperatures ($T_{\rm eff, spec}$)
  for 20 FGK members of the Hyades (orange diamonds)
  and nine of Praesepe (blue circles) are plotted
  against $T_{\rm eff, spec}$.
  $T_{\rm eff, phot}$ values are estimated based
  on the relationship between the \textit{Gaia} DR2 color \gbr\ and
  $T_{\rm eff, spec}$ for the Hyads, which we assume appear
  to us un-reddened.
  We interpret stars with $T_{\rm eff, phot} < T_{\rm eff, spec}$  as
  reddened and extinguished by interstellar dust.
  Based on Praesepe's median negative offset,
  we estimate $A_V = 0.035 \pm 0.01$ for that cluster.
    \label{f:red}}
\end{center}\end{figure}

Praesepe suffers little interstellar reddening and extinction.
Many studies---including our own prior work---assume zero reddening
\citep[e.g.,][]{douglas2014, angus2015, douglas2017, Cummings2017}
due to the cluster's close proximity to Earth.
\citet{taylor2006}, however, found $E\left(B-V\right) = 0.027$ (or $A_V = 0.084$).

Interstellar reddening is often constrained with color--color diagram or CMD analyses.
We take an alternative approach using spectroscopy. 
Co-author J.~Brewer has observed members of the Hyades and Praesepe
with Keck/HIRES for a separate project, and analyzed the spectra with Spectroscopy Made Easy \citep[][]{valenti2005} following the \citet{Brewer2015} procedure  \citep[see also][]{Brewer2016,BrewerCKS}.
We match their target list with \textit{Gaia} DR2  and filter out non-single star members according to their proximity to the empirical cluster main-sequence defined by the \citet{DR2HRD} membership list and their astrometry.
We also only focus on those stars with $5000 < \teff\ < 6200$~K, giving us 20 FGK stars in our Hyades sample and nine in our Praesepe sample.

We fit an empirical color--temperature relation to the Hyades sample, and define its reddening to be zero.
Figure \ref{f:red} compares the Praesepe
stars with their Hyades analogs,
and shows that the Praesepe stars have
photometric temperatures that are
systematically cooler than
their spectroscopic temperatures. 
Spectroscopic and photometric temperatures for individual stars are given in Table~\ref{tab:red}.
We then calculate the necessary reddening values for each star in the Hyades and Praesepe needed to align their photometric and spectroscopic temperatures.
We find $A_V = 0.035$$\pm$$0.011$
(median and 1$\sigma$) for Praesepe.
Our result splits the difference between the \citet{taylor2006} value and the oft-assumed zero reddening.

\begin{deluxetable}{lrrrr}[t]
\tablewidth{0pt}
\tabletypesize{\scriptsize}
\tablecaption{Praesepe and Hyades members used to derive the differential reddening between the two clusters \label{tab:red}}
\tablehead{
\colhead{Cluster} & \colhead{DR2Name} & \colhead{SpecTeff} & \colhead{\gbr} & \colhead{PhotTeff} 
}
\startdata
Praesepe & 662925629454594944 & 5988 & 0.785 & 5939.1342\\
Praesepe & 664683130070043136 & 5811 & 0.841 & 5758.5828\\
Praesepe & 662841379375655936 & 5420 & 0.959 & 5398.3781\\
Praesepe & 659539236719824768 & 6028 & 0.768 & 5993.4307\\
Praesepe & 659766114072052608 & 5083 & 1.111 & 5049.5066\\
Praesepe & 659343626729512832 & 6034 & 0.771 & 5981.6420\\
Praesepe & 664600804138934400 & 5390 & 0.969 & 5371.2551\\
Praesepe & 664366779961036288 & 5510 & 0.930 & 5484.3133\\
Praesepe & 659768038217395968 & 5747 & 0.858 & 5704.8021\\
Hyades & 47019347749289216 & 5141 & 1.055 & 5157.4270\\
Hyades & 52548241968465408 & 5345 & 0.979 & 5344.6499\\
Hyades & 49005581144118784 & 5527 & 0.907 & 5552.8148\\
Hyades & 47345009348203392 & 5622 & 0.886 & 5617.7784\\
Hyades & 3312644885984344704 & 5540 & 0.912 & 5535.8121\\
Hyades & 3312575685471393664 & 5938 & 0.793 & 5912.2090\\
Hyades & 3309956850635519488 & 5216 & 1.035 & 5200.8830\\
Hyades & 3309006602007842048 & 5787 & 0.831 & 5789.9270\\
Hyades & 3411887595780736128 & 5252 & 1.026 & 5223.1113\\
Hyades & 3406823245223942528 & 5273 & 1.004 & 5277.6612\\
Hyades & 3405113740864365440 & 6065 & 0.738 & 6088.7928\\
Hyades & 3407121831350730112 & 5583 & 0.894 & 5590.8317\\
Hyades & 100254161710940928 & 5787 & 0.838 & 5769.1275\\
Hyades & 8479094371605632 & 5051 & 1.107 & 5056.6689\\
Hyades & 10608573516849536 & 5959 & 0.775 & 5970.7933\\
Hyades & 149005270337201792 & 5831 & 0.821 & 5821.3471\\
Hyades & 145325548516513280 & 5498 & 0.919 & 5516.2169\\
Hyades & 3313689422030650496 & 5805 & 0.830 & 5795.3730\\
Hyades & 3306922958753764992 & 6141 & 0.725 & 6128.2910\\
Hyades & 3309170875916905856 & 5829 & 0.816 & 5840.2860\\
\enddata
\end{deluxetable}

\section{The Sun's \textit{Gaia} DR2 color}\label{sun}

Since \textit{Gaia} cannot observe the Sun's disk-integrated light,
we must instead estimate its \textit{Gaia} color with analogous field stars.
We select stars in the updated
SPOCS catalog \citep[][]{Brewer2016}
with spectroscopic properties most similar to the Sun's, identifying 11 stars with \teff\ within 100 K of 5777 K (the solar \teff\ adopted by SPOCS), $\logg > 4.3$ dex,
[Fe/H] within 0.05 dex of solar,
and $\lrphk < -4.8$ dex.

We then fit a cubic polynomial relating \teff\ to color for these stars, finding that \teff\ = 5777 K predicts a solar color ($\gbr$)$_\odot$~$= 0.817$ mag. This empirical value is in excellent agreement with that of \citet{Casagrande2018}, who estimated the solar color from a variety of spectral templates to be ($\gbr$)$_\odot$~$= 0.82$ mag.

The SPOCS star that we decided was most similar to the Sun
is HD 103828 (Gaia DR2 845471463339146496). It
has the following spectroscopic properties
in \citet{Brewer2016}:
$\teff = 5771$ K,
$\logg = 4.39$ dex,
metallicity [M/H]~= $-0.02$ dex,
and $\vsini = 1.2$ \kms.
The average chromospheric emission is
 $\lrphk =  -4.846$ \citep{Isaacson2010},
 corresponding to a chromospheric age of 3.89 Gyr \citep{mamajek2008}.
 The [Y/Mg] abundance ratio implies
 an age of 6.4 Gyr \citep{Spina2018}.
The DR2 color for this star is ($\gbr$)~$= 0.8162$ mag.

HD 222582 (Gaia DR2 2440578577126302336) is
also quite similar to the Sun, with
$\teff = 5789$ K,
$\logg = 4.38$ dex,
[M/H] = $+0.01$ dex,
and $\vsini = 0.5$ \kms.
The average chromospheric emission is
 $\lrphk =  -4.922$ \citep{Isaacson2010},
 corresponding to a chromospheric age of 5.2 Gyr \citep{mamajek2008}.
 The [Y/Mg] abundance ratio implies
 an age of 6.7 Gyr \citep{Spina2018}.
The DR2 color for this star is ($\gbr$)~$= 0.8201$~mag.

The solar twin 18 Sco (HD 146233)
has 
$\teff = 5785$ K,
$\logg = 4.41$ dex,
[M/H] = $+0.04$ dex,
$\vsini = 1.5$ \kms,
and $\lrphk =  -4.933$ dex.
It has DR2 color of 0.8081 mag,
which is only
0.009 less than our adopted solar value.

\section{Gyrochronological ages for individual Hyades and Praesepe stars}

\begin{deluxetable}{lrrrr}[h]
\tablewidth{0pt}
\tabletypesize{\scriptsize}
\tablecaption{gyrochronological ages for cluster members \label{tab:ages}}
\tablehead{
\colhead{Cluster} & \colhead{DR2Name} & \colhead{EPIC} & \colhead{PhotTeff} & \colhead{Gyro Age}
\\
\colhead{} & \colhead{} & \colhead{} & \colhead{(K)} & \colhead{(Myr)}
}
\startdata
Praesepe & 661311752544248960 & 211971871 & 6289.11 & 634.714\\
Praesepe & 661401122222444032 & 211980688 & 6014.5 & 706.947\\
Praesepe & 661317250102375040 & 211974702 & 5495.69 & 620.683\\
Praesepe & 661319483485360000 & 211979334 & 5832.3 & 639.12\\
Praesepe & 661243273585808000 & 211949471 & 6042.46 & 739.369\\
Praesepe & 661277461525419008 & 211947686 & 6054.86 & 657.916\\
Praesepe & 661319547906689024 & 211980170 & 5204.55 & 644.999\\
Praesepe & 661292511090872960 & 211956059 & 5659.8 & 666.701\\
Praesepe & 661244270018200064 & 211952381 & 6333.42 & 934.809\\
Praesepe & 661207024061874944 & 211926132 & 6098.54 & 670.077\\
Praesepe & 664387808120781056 & 211995288 & 5678.54 & 602.454\\
Praesepe & 664302424170984960 & 211971690 & 5684.81 & 740.524\\
Praesepe & 661424074527577984 & 211998346 & 6117.38 & 721.674\\
Praesepe & 659687498990893056 & 211910082 & 6026.83 & 672.431\\
Praesepe & 664403991557399040 & 212018902 & 5584.4 & 687.17\\
Praesepe & 659758967246507008 & 211934056 & 5391.99 & 623.933\\
Praesepe & 664311392062658048 & 211983461 & 5740.2 & 653.706\\
Praesepe & 664437286143574016 & 212012299 & 5078.86 & 677.003\\
Praesepe & 661028662658248960 & 211925093 & 5469.56 & 663.474\\
Praesepe & 661401225301656064 & 211982334 & 5449.49 & 734.842\\
Praesepe & 661300993647980032 & 211967293 & 5181.36 & 656.705\\
Praesepe & 664334035130360064 & 211983499 & 5037.35 & 611.489\\
Praesepe & 660998975844267008 & 211911846 & 5554.36 & 693.469\\
Praesepe & 664283079638402944 & 211959522 & 5405.09 & 630.049\\
Praesepe & 659755630055476992 & 211925552 & 5075.03 & 639.126\\
Praesepe & 664366779961036032 & 211992034 & 5494.28 & 695.63\\
Praesepe & 661029384212747008 & 211929531 & 5208.1 & 734.605\\
Praesepe & 664497209527232000 & 212019439 & 5058.83 & 620.356\\
Praesepe & 659665096441437056 & 211895099 & 6016.77 & 890.296\\
Praesepe & 665129291274350976 & 212075775 & 5492.58 & 674.2\\
Praesepe & 661422837576999040 & 211994672 & 5228.23 & 704.099\\
Praesepe & 661295324291154944 & 211959779 & 5309.35 & 532.333\\
Praesepe & 661029727808614016 & 211927269 & 5230.53 & 649.736\\
Praesepe & 659766114072052992 & 211931128 & 5026.11 & 687.499\\
Praesepe & 661338858079664000 & 211967873 & 5035.94 & 680.82\\
Praesepe & 660944717521395968 & 211900700 & 5105.37 & 827.693\\
Praesepe & 661222279785744000 & 211936827 & 5167.1 & 512.48\\
Praesepe & 665004702861616000 & 212080687 & 6164.63 & 667.983\\
Praesepe & 664283522018091008 & 211958260 & 5060.73 & 658.382\\
Praesepe & 659343626729512960 & 211842439 & 5971.6 & 698.018\\
Praesepe & 661325908755276032 & 211953567 & 5470.68 & 669.828\\
Praesepe & 661239390935361024 & 211948267 & 5023.77 & 660.063\\
Praesepe & 662925629454594944 & 211955365 & 5929.1 & 622.924\\
Hyades & 3313662896312488192 &  & 5896.61 & 756.949\\
Hyades & 3314109916508904064 &  & 5917.49 & 708.125\\
Hyades & 45367056650753280 &  & 5954.75 & 767.84\\
Hyades & 3312575685471393664 &  & 5856.73 & 885.692\\
Hyades & 149005270337201792 & 211037886 & 5769.26 & 726.991\\
Hyades & 3313689422030650496 &  & 5744.58 & 752.555\\
Hyades & 3309006602007842048 &  & 5739.42 & 650.624\\
Hyades & 47345009348203392 &  & 5578.14 & 710.055\\
Hyades & 48061409893621248 &  & 5513.63 & 1029.32\\
Hyades & 49005581144118784 &  & 5517.61 & 745.346\\
Hyades & 3312644885984344704 &  & 5501.74 & 793.994\\
Hyades & 145325548516513280 & 210899260 & 5483.42 & 703.103\\
Hyades & 68001499939741440 &  & 5213.03 & 676.919\\
Hyades & 3314213025787054592 & 210666330 & 5286.06 & 851.798\\
Hyades & 43538293935879680 &  & 5090.25 & 700.693\\
\enddata
\end{deluxetable}


\vspace{5mm}
\facilities{Kepler (K2)}

\software{Astropy \citep{astropy},
        Astroquery \citep{ginsburg2013},
        AstroML \citep{vanderplas2012,ivezic2013},
        pywcsgrid2 (J.~Lee),\footnote{\url{https://github.com/leejjoon/pywcsgrid2}}
        K2fov \citep{mullally2016}}


\setlength{\baselineskip}{0.6\baselineskip}
\bibliography{references}
\setlength{\baselineskip}{1.667\baselineskip}

\end{document}